\documentclass[sigconf, nonacm]{acmart} 

\usepackage{graphicx}
\usepackage{subcaption}
\usepackage{adjustbox}
\usepackage{xspace}

\AtBeginDocument{}

\DeclareRobustCommand{\pumpfun}{\texttt{pump.fun}\xspace}
\DeclareRobustCommand{\onepctsample}{\textbf{1\% coin sample}\xspace}
\DeclareRobustCommand{\fivedaysample}{\textbf{5-day sample}\xspace}
\DeclareRobustCommand{\artifactURL}{\url{https://www.doi.org/10.1184/R1/33420628}}
\begin{document}

\title[Meme Coin Factories: Uncovering Large-Scale Manipulations on \pumpfun]{Meme Coin Factories:\\ Uncovering Large-Scale Manipulations on \pumpfun} 

\author{Nicolas Szwajcok$^{*,1,2,3}$, Taro Tsuchiya$^{*,3}$, Enze Liu$^{3}$, 
Kyle Soska$^4$, Mathias Payer$^1$, Nicolas Christin$^3$
\\
$^1$ \'{E}cole Polytechnique F\'{e}d\'{e}rale de Lausanne\\
$^2$ ETH Z\"{u}rich\\
$^3$ Carnegie Mellon University\\
$^4$ Fullstack
}
\thanks{$^*$ Joint first authors with equal contribution, ordered alphabetically.}

\renewcommand{\shortauthors}{Szwajcok, Tsuchiya et al.}

\begin{abstract} 
Once complex, creating and deploying a new cryptocurrency has become trivial. 
\emph{Coin launchpads} now allow users to generate a new coin with merely a few clicks, at a minimal cost.
Launchpad popularity has grown in tandem with the rise of ``meme coins,'' which usually do not offer any novel technological properties and
are purely created for fun. 
The most prominent coin launchpad, \pumpfun, has gained significant traction, grossing over 100~million USD in daily trading volume. 
The mass adoption of coin launchpads, however, also enables strategic actors to easily manipulate trading signals, unbeknownst to inexperienced traders who then buy certain coins, and enable these strategic actors to profit from rapid and unsustainable price increases (``pumps''). 
To identify such manipulations at scale, we conduct a large-scale study of \pumpfun, collecting information on all 15 million coins launched in the last two years, and performing analysis on large, random samples of transaction data.
We identify five classes of manipulation strategies: 1) wash trading,
2) creator address obfuscation,
3) coordinated sell, 
4) copycat coins,
and 5) social media manipulation.
We find that strategic actors often bypass the platform interface and implement these strategies in a highly automated and low-latency fashion, by interacting directly with the blockchain.
We further uncover the existence of ``Market-Manipulation-as-a-service (MMaaS),'' third-party tools that enable users to perform these manipulations without any technical expertise. 
We conclude by devising mitigations and proposing recommendations for traders, \pumpfun, wallets or chain scanners, software development platforms,
and regulators.\end{abstract}



\maketitle

\section{Introduction} 
\label{sec:intro}
When Bitcoin rose to prominence in the early 2010s,
competitors, or ``alt-coins,'' started to appear. At the time, creating
an alt-coin required technical expertise to develop and
deploy a new blockchain, or ``forking'' an existing one. 
Subsequently, the rise of programmable money, in particular through Ethereum's
smart contracts, made alt-coin creation significantly less onerous.
Ethereum specifies a well-defined ERC-20 standard to allow users to create ``tokens'' on an existing blockchain as a substrate. 
The mid-2020s made token creation even
easier, with the introduction of online platforms and mobile apps---called \emph{token
launchpads}---making the generation of a new token 
literally as simple as pushing a button on a website. 
Token launchpads often operate on the Solana blockchain, which supports low transaction fees and high throughput, making coin creation nearly zero-cost.
These tokens are often referred to as ``meme coins.\footnote{By a slight abuse of
terminology, we will indistinctly refer to these as \emph{tokens} or
as \emph{coins}. Technically, a coin requires to operate its
own blockchain, and we should denote them as ``meme tokens,'' but
``meme coin'' is the more common turn of phrase.}'' 
They generally do not offer any novel technological properties, and are instead mostly designed to be humorous~\cite{tanner2025who_gets_to_have_fun}.

\pumpfun has been the most prominent token launchpad, attracting over 15 million new coins and totaling nearly 90 billion USD in trading volume~\cite{dune2026pumpfun,blockworks2026pumpfun}.                               
For instance, the most popular coin, ``Fartcoin,'' reached a market capitalization of 2.5 billion USD at its peak.
Importantly, \pumpfun is extremely user-friendly.
To create a coin, users only need to provide a coin name, symbol, coin image, and \pumpfun automatically handles the technical details such as  instantiating tokens on the Solana blockchain and generating a trading pool. 
This ease of use, combined with potential financial gains, attracts users with no technical expertise, potentially including teenagers~\cite{BoumaSims:SOUPS24}.
Additionally, users only need to connect their Solana address
and can remain pseudonymous. No identity verification or any other
information---not even an email address---is required.

The user-friendliness and openness, however, create opportunities for \emph{strategic actors} to take advantage of inexperienced traders.
In particular, strategic actors often manipulate trading signals (e.g., trading volume, ownership, coin names, social media comments, and posts) 
to trick inexperienced traders into buying certain coins and profit from the resulting price increase. 
Inexperienced traders often fail to detect such highly automated, low-latency manipulations, which are buried in massive trading volume.

This study aims to detect market manipulation on \pumpfun at scale and uncover manipulation strategies and infrastructure. 
We focus on \pumpfun, because its trading volume far surpasses that of all its competitors (e.g., BONKfun, Raydium LaunchLab), which together feature only about 1~million coins and a combined market cap of 7~billion USD~\cite{blockworks2026raydium}, compared to 15.2 million coins on \pumpfun alone. 
We track \textit{all} 15.2 million \pumpfun coins on the Solana blockchain and collect their metadata (e.g., coin description, image URLs, social media links) during the first two years of \pumpfun's existence (Jan. 14, 2024 -- Jan. 14, 2026). 
For \S\ref{wash_trading_section}-\ref{sec:evaluation_dumping}, we collect the full transaction history of a random \onepctsample (as the primary sample) and all coins launched on a \fivedaysample (for consistency checks), totaling over 87 million transactions.

\begin{figure}
    \centering
    \includegraphics[width=\linewidth]
    {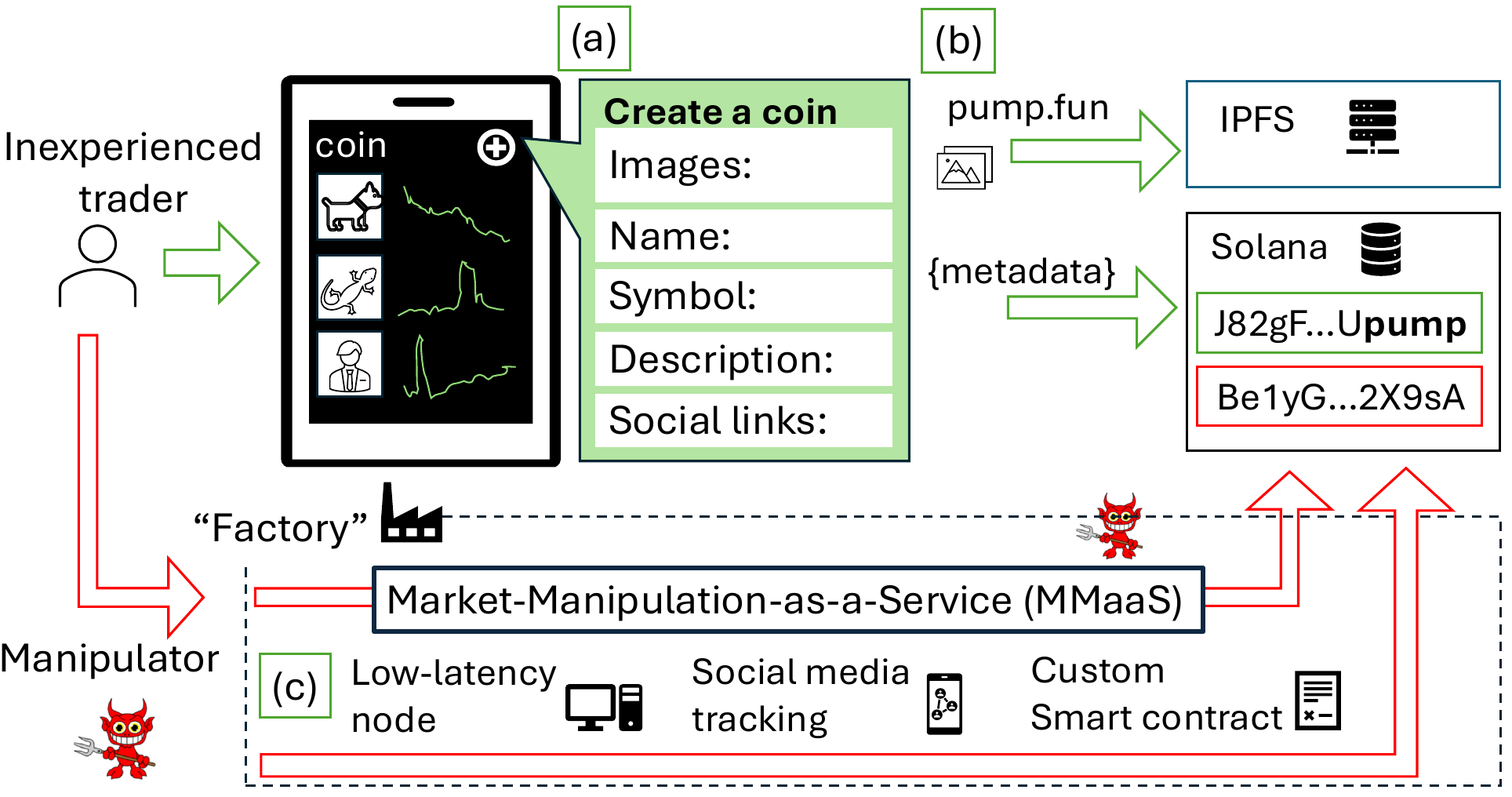}
    \caption{How to create a coin on \pumpfun (both for inexperienced traders and strategic actors)}
\label{fig:manipulation_overview}
\end{figure}

We identify five classes of manipulation strategies:

\noindent\textbf{1. Wash trading.} 
Manipulators can perform buy-sell transactions to inflate trading volumes (``wash trading'').
We identify a lower-bound of 4 million wash trading transactions (2.2 million for the 1\% coin sample, 1.8 million for the 5-day sample), representing 17\% of all \emph{trading} transactions.
We show that wash trading can potentially increase a coin success probability (i.e., ``graduation'' defined in \S\ref{subsec:bg_pumpfun}) up to 10 times.

\noindent \textbf{2. Creator address obfuscation.} 
Manipulators can use a different blockchain address to create coins, potentially hiding their coin ownership. 
We develop a method to group coin creator addresses based on funding patterns (e.g., which address funds a coin creator address).
We show that the top-1\% groups create 58.6\% of all coins.

\noindent \textbf{3. Coordinated sell.} 
Manipulators can use multiple blockchain addresses to buy tokens and consolidate them into a single address, which then sells them all at once (``dump'').
Using conservative criteria, we identify nearly 8,000 instances (4,402 for the 1\% coin sample and 3,510 for the five-day sample) of such manipulations.

\noindent \textbf{4. Copycat coins.} 
Manipulators often deploy ``copycat'' coins, replicas of the existing coins, to lure uninformed traders.
We detect at least 1.5 million (over 10\% of all) coins that duplicate the name, symbol, description, and images of original coins. 
We find that manipulators often automate copycat creation (e.g., through a custom smart contract), while the original coins are much more successful. 

\noindent \textbf{5. Social media manipulation.} 
We identify and analyze social media manipulation, including bots posting promotional comments on a coin profile pages, in Telegram groups, and in Twitter communities that consistently generate hundreds of coins. 
We also show that 3.5 million (23.5\% of all) coins are created after Twitter or Truth Social posts.
Among these, we observe that 31 posts \emph{each} helped their respective coin creators earn \emph{at least} 1 million USD.

Across all these strategies, we find evidence of sophistication.
As shown in Figure~\ref{fig:manipulation_overview}, strategic actors often do not rely on the \pumpfun web interface, but instead directly interact with the Solana blockchain, most likely using the custom programs (e.g., smart contracts) and low-latency blockchain nodes: ``Meme Coin Factories.''
We also show that some of these strategies are effective at raising prices and extracting financial value. 

As conducting such manipulations requires technical expertise and infrastructure, we further identify the existence of  ``Market-Manipulation-as-a-service (MMaaS),'' websites or software tools that help commoditize such manipulations at scale. 
Our qualitative analysis reveals that MMaaS tools advertise the ability to create multiple blockchain addresses, copy tokens with low latency, generate comments using custom templates or AI, and deploy anti-detection tools (e.g., CAPTCHA solvers or proxy-rotation).

Compared to contemporary studies, our core contribution is twofold (see \S\ref{subsec:related_work} for more details).
First, we analyze known manipulation strategies (e.g., wash trading, dump) 
on a substantially larger scale (and using uniform samples) than has been studied before, thereby illuminating 
the pervasiveness of these activities. 
Second, we identify novel manipulative strategies (e.g., social media \textit{post} manipulation, MMaaS). 
Based on these additional insights, we propose specific 1) improvements to trade signals (e.g., for users, \pumpfun, wallets, and chain scanners) and 2) interventions (e.g., for financial regulators and software development platforms).

 \section{Background}
\label{sec:bg}
We introduce \pumpfun, describe our threat model, and highlight our contributions compared to related work.

\subsection{\pumpfun}
\label{subsec:bg_pumpfun}
Figure~\ref{fig:manipulation_overview} illustrates the coin creation process on \pumpfun. 
When users create a coin on \pumpfun, they provide a coin name, symbol, and image.
They can optionally provide a coin description and links to external platforms, e.g., Twitter or Telegram (Figure~\ref{fig:manipulation_overview} (a)).
After \pumpfun receives this information, it uploads a coin image to the IPFS distributed file sharing service, 
records this information on the Solana blockchain (in ``programs'' similar to smart contracts in Ethereum), and generates a coin address ending with ``pump'' (Figure~\ref{fig:manipulation_overview} (b)). 
\pumpfun keeps track of the coin information through the \pumpfun's Token Mint Authority address.\footnote{TSLvdd1pWpHVjahSpsvCXUbgwsL3JAcvokwaKt1eokM} Simultaneously, \pumpfun creates a trading pool called ``bonding curve'' and inserts 1 billion coin tokens into it. 
Users send SOL (i.e., the native Solana cryptocurrency) to the pool to buy tokens, and can, conversely, redeem SOL from the pool to sell tokens.
After the first trade, the trading pool uses an automated market maker (AMM) algorithm to determine the price based on the amount of SOL and the coin tokens in the pool.
Sophisticated users can bypass the \pumpfun web interface and perform the above actions directly on Solana through custom smart contracts (Figure~\ref{fig:manipulation_overview} (c)).  

When a bonding curve retains about 85 SOL, \pumpfun moves the coin to an external decentralized exchange (``DEX,'' such as Raydium and PumpSwap) and stops trading on the launchpad~\cite{bitquery_2026_understanding_pump_fun_doc,pumpfun2025fees_graduation}.
\pumpfun calls this process ``\emph{graduation}'' and considers it a success indicator that attracts greater public visibility.  
Unlike the bonding curve, a DEX has liquidity providers (who deposit both SOL and tokens to earn rewards).  
In our sample, 1.02\% of coins graduate.

\subsection{Threat model}
We define a manipulator as a strategic actor who engages in the following process.
Manipulators first fabricate signals of coin demand, such as trading signals (e.g., volume) or social signals (e.g., comments), or impersonate existing coins. 
The goal is to get a (set of) target coin(s) to the top of the \pumpfun webpage to advertise it~\cite{luo2026resisting, kawai2024stranger}. 
This bump leads inexperienced traders to buy the target coin(s). 
The manipulator can then profit from this demand by selling previously accumulated target coin(s) at higher prices.

Manipulators can interact directly with the Solana blockchain via custom smart contracts, rather than manually using the \pumpfun web interface. 
For instance, they can bundle many trades \textit{within} a single transaction. 
We also assume that manipulators can create many Solana addresses, access low-latency Solana nodes to execute trades faster, and track social media posts in real time using platform APIs.
We do not assume that manipulators control the blockchain network, compromise users' private keys, or control the \pumpfun platform itself.
Manipulators follow existing protocol rules and perform standard Solana transactions.

On the other hand, we define an ``inexperienced'' trader as a user who primarily uses the \pumpfun's web interface to create or trade coins. 
They can use wallet services and blockchain explorers to gather coin information (e.g., owner holdings) or browse social media for related posts. 
However, we do not assume that inexperienced traders possess sufficient technical expertise to interact with or analyze data on the Solana blockchain, or to programmatically retrieve information from social media (e.g., via APIs).

\subsection{Manipulation strategies}
\label{subsec:bg_manipulation_strategies}
This section briefly introduces the five manipulative strategies and MMaaS, and surveys the existing literature on each strategy.

\noindent \textbf{1. Wash trading.}
Wash trading is a manipulation to generate artificial demand by selling and buying assets to oneself, while leaving the asset balance unchanged~\cite{doj_wash_trades_press_release, securities_exchange_act_78i, cftc_wash_trading_glossary}.
Manipulators can also use multiple accounts to obfuscate their activity while keeping their overall balance unchanged. 

\noindent \textbf{2. Creator  address obfuscation.}
Traders often consider a coin unsustainable if a coin creator or a single user owns a disproportionately large share of the circulating supply. Manipulators can conceal this by using a different blockchain address to create a coin.

\noindent \textbf{3. Coordinated sell.}
Pump-and-dump is a trading scheme in which a group of traders coordinates to buy a certain asset, often advertising on social media, and then sells it at a higher price for profit~\cite{li2025cryptocurrency}. 
In particular, manipulators can create multiple accounts to buy tokens, transfer them to a single account, and sell them at once.

\noindent \textbf{4. Copycat coins.}
Manipulators often impersonate an authentic asset and create counterfeit ones (e.g., a slightly different symbol) and trick victims into buying them for profit~\cite{gao2020tracking, xia2021trade, cernera2023token}. 
Most blockchains (including Solana) have no restrictions on naming coins; users can copy-paste the existing coin information (e.g., name, symbol, description, image) to create a replica, which  we call ``copycat'' coins.
On \pumpfun, a manipulator, as a coin creator, receives shares from trades victims make (i.e., a coin creator fee) and profits from price increases in copycat coins, offset by coin-creation fees.

\noindent \textbf{5. Social media manipulation.}
Users often seek social signals (e.g., comments about coins) to look for a potential price increase~\cite{xu2019anatomy,mirtaheri2021identifying,nizzoli2020charting,tsuchiya2021profitability}. 
Manipulators can deploy automated bots to post comments on the coin's profile page on \pumpfun, organize social media groups and posts to create a false sense of market excitement. 
Furthermore, social media influencers can create a Twitter post and a coin simultaneously to extract profits from the price increase.

\noindent \textbf{MMaaS.}
Performing market manipulation is daunting for non-tech-savvy users because it requires technical expertise (e.g., creating blockchain addresses and smart contracts) and access to lower-latency blockchain nodes to achieve better profits. 
We observe a surge in trading apps, websites, and software tools that offer a user-friendly interface for executing trades on behalf of users, typically charging per-trade fees (e.g., 1\%).  
Alarmingly, some apps could facilitate various forms of market manipulations. 
The US SEC describes such services as ``Market-Manipulation-as-a-Service'' (MMaaS, \cite{sec2024mmaas}); one court case already exists~\cite{masssachusetts2024mytrade}. 
This phenomenon mirrors broader cybercrime-as-a-service, including spam-as-a-service~\cite{thomas2011suspended}, ransomware-as-a-service~\cite{cong2025anatomy}, DDoS-for-hire services~\cite{vu2025assessing}, and front-running-as-a-service~\cite{felton2020front}.

\subsection{Related work}
\label{subsec:related_work}
There exists an extensive line of work on pump-and-dumps~\cite{kramer2005way, li2025cryptocurrency, kamps2018moon, la2023doge, victor2019cryptocurrency, hamrick2021examination, dhawan2023new}, wash trading~\cite{cong2023crypto, falk2023can, mongardini2025midsummer}, and counterfeit tokens~\cite{gao2020tracking, xia2021trade, cernera2023token, ye2024interface, guan2024characterizing, tsuchiya2025address}, both in traditional and cryptocurrency markets. 
Most of these studies focus on centralized exchanges and rely on aggregated market data, such as coin prices and volumes, to infer manipulations.
With the rise of high-throughput, low-fee blockchains, strategic actors began to execute their manipulations on the blockchain directly.
A handful of studies started analyzing manipulation at account-level~\cite{victor2021detecting,falk2023can,gan2024exposing}.
We further identify the use of \textit{multiple} trading accounts within a single manipulation (in \S\ref{sec:evaluation_dumping}), which is challenging due to data availability and complexity. 

While there also exist a few studies of meme coins~\cite{long2025bridging, li2025trust, mongardini2025midsummer, luo2026resisting}, the work closest to ours is Ding et al.~\cite{ding2025decompose}, who 
study wash trading, owner obfuscation, and comment-generating bots on \pumpfun and investigate their impact on coin outcome (e.g., price change). 

Our work differs from previous efforts as follows. 
First, our paper draws conclusions from uniformly sampled and significantly larger datasets.  
We analyze metadata for \emph{all} 15.2 million coins and the full transaction data for over 300,000 coins,  
whereas Ding et al.~\cite{ding2025decompose} analyze only about 6,000 coins (i.e., 0.04\% of all coins) before and after the TRUMP coin. 
Our copycat coin analysis (1.5 million) is also on a much larger scale compared to other counterfeit token studies mentioned above.  
Second, we introduce more granular detection methods and analyses. 
For instance, Ding et al.~\cite{ding2025decompose} identified wash trading \textit{accounts} that repeatedly buy and sell tokens (by setting thresholds).
Instead, our work directly identifies wash trading \textit{events}. 
This approach allows us to infer manipulators’ intent and capability better and produces fewer false positives (i.e., WT1, \S\ref{wash_trading_section}).
Last, our paper is the first to analyze creator obfuscation (using different addresses to create coins), social media \textit{post} manipulation (making a post that creates new coins), and MMaaS.

 \newcommand{\NSelectedDays}{five\xspace}
\newcommand{\RandomSampleRate}{1\%\xspace}
\newcommand{\NMintTx}{15,246,041\xspace}
\newcommand{\NUniqueMints}{15,245,966\xspace}
\newcommand{\NCandidateSignatures}{24,577,488\xspace} \newcommand{\NFilteredSignatures}{9,331,447\xspace}

\newcommand{\NDays}{730\xspace}
\newcommand{\NSlots}{151,780,482\xspace}
\newcommand{\MinSlot}{241,768,575\xspace}
\newcommand{\MaxSlot}{393,549,057}

\newcommand{\MedianDailyMints}{17,926\xspace}
\newcommand{\MaxDailyMints}{71,735\xspace}

\newcommand{\NUniqueCreators}{5,826,346\xspace}
\newcommand{\PctNonNullName}{99.71\%\xspace}
\newcommand{\PctNonNullSymbol}{99.68\%\xspace}
\newcommand{\PctNonNullURI}{99.75\%\xspace}
\newcommand{\PctCreatorsMultiMint}{21.67\%\xspace}
\newcommand{\MedianMintsPerCreator}{1\xspace}
\newcommand{\MaxMintsPerCreator}{40,760\xspace}

\newcommand{\MedianMintingFeeSol}{0.00033\xspace}
\newcommand{\MedianMintingFeeUsd}{0.03}

\section{Datasets}
\label{sec:dataset}
We collect both on-chain and external data related to coins created on \pumpfun. We first describe our on-chain data collection and then outline the external data sources we use to enrich the analysis.

\subsection{On-chain Data}
Our on-chain dataset captures the creation and trading of coins launched on \pumpfun. 
We obtain this data from three third-party providers: Chainstack, Syndica, and Arkham.

\subsubsection{Coin addresses and creators}
\label{sec:dataset:minting}
We begin by identifying all coins created on \pumpfun in its first two years of existence (Jan. 14, 2024--Jan. 14, 2026). 
Conveniently, \pumpfun uses a single address, the \pumpfun Token Mint Authority, to track coins (see \S\ref{subsec:bg_pumpfun}).
We retrieve all transactions involving this address and filter for transactions that are parsable and successfully create new coins.

\begin{figure}[t]
\centering
\includegraphics[width=\columnwidth]{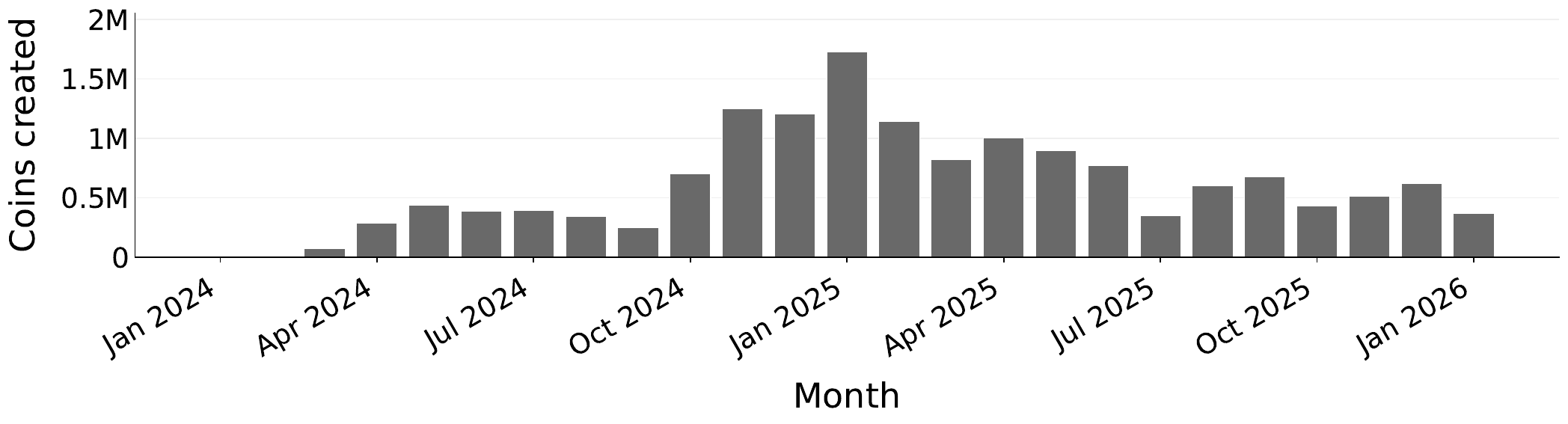}
\caption{Nr. of \pumpfun coins for each month.}
\label{all_coins_created}

\end{figure}

In total, we obtain \NUniqueMints coins. 
This dataset spans \NDays days and \NSlots Solana blockchain slots (\MinSlot to \MaxSlot). 
Figure~\ref{all_coins_created} illustrates the distribution of coins created over time. 
\pumpfun has remained active during the data collection period, with a median of \MedianDailyMints new coins per day and a peak of \MaxDailyMints new coins on a single day.

We extract the \emph{creator address} for each coin, defined as the address that pays the transaction fees for the first transaction.
We also capture the \emph{funding addresses} of each creator address. This set of addresses consists of the \emph{funder address}, that is, the first address to send any SOL or any tokens to the creator; the parent of the funder address (the first address to have sent anything to the funder); and its own parent. 
In other words, 
we trace the funding addresses up to three hops (addresses) backward. 
We empirically decided on a three-hop threshold, based on saturation in new addresses at that level (see details in \S\ref{sec:evaluation_clustering}).
In total, we obtain \NUniqueCreators distinct creator addresses and their associated funding addresses.

\subsubsection{Sampled coin transactions}
\label{sec:dataset:transactions}

\newcommand{\MaxSampleTxPerToken}{5,000,000\xspace}

\newcommand{\NSelectedDayTokens}{149,028\xspace}
\newcommand{\NSelectedDayTx}{37,350,061\xspace}
\newcommand{\NSelectedDayPumpEvents}{31,078,502\xspace}
\newcommand{\NSelectedDayTransfers}{26,537,567\xspace}

\newcommand{\NSelectedDayRawTokensBeforeExcluding}{149,038\xspace}
\newcommand{\NSelectedDayExcludedHeavyTokens}{10\xspace}
\newcommand{\PctSelectedDayExcludedHeavyTokens}{0.007\%}

\newcommand{\NRandomEligibleTokens}{152,176\xspace}
\newcommand{\NRandomExcludedHeavyTokens}{5\xspace}
\newcommand{\PctRandomExcludedHeavyTokens}{0.003\%}
\newcommand{\NRandomSampleTokens}{152,171\xspace}
\newcommand{\NRandomSampleTx}{49,970,921\xspace}
\newcommand{\NRandomSamplePumpEvents}{31,078,502\xspace}
\newcommand{\NRandomSampleTransfers}{33,120,555\xspace}

\newcommand{\NAllHistoryTokens}{299,799\xspace}
\newcommand{\NAllHistoryTx}{87,092,997\xspace}
Due to the extremely large volume of coins and transactions, it is cost-prohibitive to retrieve and analyze all transactions for all coins. 
We therefore randomly sample 1\% of coins out of all coins created (\onepctsample) to reduce sampling bias. 
We also create a second sample that includes coins created on \NSelectedDays selected days (\fivedaysample) to demonstrate that trends are consistent across samples.
We exclude any coin, in either sample, that features more than 5~million transactions, as including those would triple the size of our database.
For the remaining coins, we retrieve all transactions involving these coins and conduct a detailed analysis in \S\ref{wash_trading_section} and \S\ref{sec:evaluation_dumping}.

\noindent \onepctsample.
The first sample is built from a 1\% random coin sample. After excluding 5 coins with more than 5 million transactions, the sample contains \NRandomSampleTokens coins and \NRandomSampleTx transactions.

\noindent \fivedaysample.
The second dataset includes coins created on \NSelectedDays days: Jan. 19, 2025, Aug. 1, 2025, Oct. 22, 2024, Mar. 13, 2025, and Sep. 3, 2025. 
Jan. 19, 2025 and Aug. 1, 2025 had exceptionally high and low coin creation activity, while Oct. 22, 2024, Mar. 13, 2025, and Sep. 3, 2025 were chosen at random. 
After excluding 10 coins, the sample contains \NSelectedDayTokens coins and \NSelectedDayTx transactions.

\newcommand{\NArkhamAddresses}{7,154,746\xspace}

\subsection{External Data}
\label{subsec:external_data}
We augment the on-chain data with external data sources, including address labels using the Arkham API~\cite{arkham_api_guide}, and data from the \pumpfun API~\cite{bankk2026pumpfun_api}. 
We followed the Terms of Use of \pumpfun and maintained 4--5 second intervals between requests.

\subsubsection{Address labels}
\label{sec:dataset:arkham}
Arkham labels---applied to creator and funding addresses---contain useful information, such as whether the address belongs to an exchange. 
These labels allow us to identify the types of entities funding the coin creators, aid our analysis on funding relationships, and address obfuscation (\S\ref{sec:evaluation_clustering}).

\subsubsection{Data from \pumpfun}
\label{subsec:off-chain_data}
Data from \pumpfun consists of coin metadata, user comments, and historical price data.

\noindent \textbf{Coin metadata.} 
We collect the following metadata for all \NUniqueMints coins in our on-chain dataset: 
1) basic information, such as coin name, symbol, description, and creation timestamp, 2) metrics maintained by \pumpfun, such as whether the coin has graduated and the number of comments in the coin profile page, and 3) social media links (e.g., Twitter), and links to the coin's profile images. 
In total, we retrieve metadata for 15,183,009 coins (99.59\%).  

\noindent \textbf{User comments for selected coins.} We collect comments that users post on each coin's profile page on \pumpfun to analyze potential social signal manipulations. 
Due to rate limits, we only collect comments for coins in the \onepctsample and filter out coins with no comments. 
58,715 (out of \NRandomSampleTokens) coins have at least one comment, totaling 694,483 comments.

\noindent \textbf{Historical price data for graduated coins.} 
We retrieve historical daily price data (open, close, high, low) for all graduated coins to estimate potential profits from manipulations.
We only gather price data for graduated coins because non-graduated coins are economically less significant.
In total, we have price data for 154,513 graduated coins, with 57,623,422 daily price data points.

 \newcommand{\NSelectedHOneWashTokens}{9,027 }
\newcommand{\PctSelectedHOneWashTokens}{6.07\% }

\newcommand{\NSelectedHTwoWashTokens}{132,305 }
\newcommand{\PctSelectedHTwoWashTokens}{89.01\% }

\newcommand{\NSelectedHThreeWashTokens}{139,856 }
\newcommand{\PctSelectedHThreeWashTokens}{94.09\% }

\newcommand{\NRandomHOneWashTokens}{12,495 }
\newcommand{\PctRandomHOneWashTokens}{8.26\% }

\newcommand{\NRandomHTwoWashTokens}{135,240 }
\newcommand{\PctRandomHTwoWashTokens}{89.37\% }

\newcommand{\NRandomHThreeWashTokens}{144,318 }
\newcommand{\PctRandomHThreeWashTokens}{95.37\% }

\newcommand{\SelectedHOneWashVolumeSol}{51,792 } \newcommand{\PctSelectedHOneWashVolume}{1.05\% }

\newcommand{\SelectedHTwoWashVolumeSol}{748,538 } \newcommand{\PctSelectedHTwoWashVolume}{15.15\%}

\newcommand{\SelectedHThreeWashVolumeSol}{3,077,045 } \newcommand{\PctSelectedHThreeWashVolume}{62.27\%}

\newcommand{\PctSelectedHOneMedianWashShare}{2.52\% }
\newcommand{\PctSelectedHOneMedianNoWashShare}{0.00\% }

\newcommand{\RandomHOneWashVolumeSol}{66,636 } \newcommand{\PctRandomHOneWashVolume}{1.32\%\xspace}

\newcommand{\RandomHTwoWashVolumeSol}{709,900 } \newcommand{\PctRandomHTwoWashVolume}{14.05\%}

\newcommand{\RandomHThreeWashVolumeSol}{2,950,377 } \newcommand{\PctRandomHThreeWashVolume}{58.41\%}

\newcommand{\PctRandomHOneMedianWashShare}{1.00\% }
\newcommand{\PctRandomHOneMedianNoWashShare}{0.00\%}

\newcommand{\PctSelectedHOneWashGraduation}{2.81\% }
\newcommand{\PctSelectedNoHOneWashGraduation}{1.13\% }

\newcommand{\PctSelectedHTwoWashGraduation}{1.13\% }
\newcommand{\PctSelectedNoHTwoWashGraduation}{2.09\% }

\newcommand{\PctSelectedHThreeWashGraduation}{1.19\% }
\newcommand{\PctSelectedNoHThreeWashGraduation}{1.89\% }

\newcommand{\NSelectedNoHOneWashTokens}{139,614 }
\newcommand{\NSelectedNoHTwoWashTokens}{16,336 }
\newcommand{\NSelectedNoHThreeWashTokens}{8,785 }

\newcommand{\PctRandomHOneWashGraduation}{2.00\% }
\newcommand{\PctRandomNoHOneWashGraduation}{0.90\%}

\newcommand{\PctRandomHTwoWashGraduation}{0.90\% }
\newcommand{\PctRandomNoHTwoWashGraduation}{1.76\% }

\newcommand{\PctRandomHThreeWashGraduation}{0.94\% }
\newcommand{\PctRandomNoHThreeWashGraduation}{1.92\%}

\newcommand{\NRandomNoHOneWashTokens}{138,839 }
\newcommand{\NRandomNoHTwoWashTokens}{16,094 }
\newcommand{\NRandomNoHThreeWashTokens}{7,016 }

\newcommand{\NSelectedNoWashTokens}{8,774 }
\newcommand{\PctSelectedNoWashTokens}{5.91\% }

\newcommand{\NRandomNoWashTokens}{7,016 }
\newcommand{\PctRandomNoWashTokens}{4.63\% }

\newcommand{\PctRandomGradCohortZero}{0.90\% }
\newcommand{\NRandomGradCohortZero}{138,839 }

\newcommand{\PctRandomGradCohortOneToFive}{0.50\% }
\newcommand{\NRandomGradCohortOneToFive}{2,618 }

\newcommand{\PctRandomGradCohortSixToTwentyFive}{0.65\% }
\newcommand{\NRandomGradCohortSixToTwentyFive}{3,873 }

\newcommand{\PctRandomGradCohortTwentySixToHundred}{1.12\% }
\newcommand{\NRandomGradCohortTwentySixToHundred}{2,582 }

\newcommand{\PctRandomGradCohortHundredOneToFiveHundred}{3.61\% }
\newcommand{\NRandomGradCohortHundredOneToFiveHundred}{2,437 }

\newcommand{\PctRandomGradCohortFiveHundredPlus}{9.64\% }
\newcommand{\NRandomGradCohortFiveHundredPlus}{985 }

\newcommand{\PctRandomMeanWashShareTxOneToTen}{0.41\% }
\newcommand{\NRandomMeanWashShareTxOneToTen}{58,084 }

\newcommand{\PctRandomMeanWashShareTxElevenToHundred}{2.69\% }
\newcommand{\NRandomMeanWashShareTxElevenToHundred}{75,023 }

\newcommand{\PctRandomMeanWashShareTxHundredOneToThousand}{13.91\% }
\newcommand{\NRandomMeanWashShareTxHundredOneToThousand}{15,905 }

\newcommand{\PctRandomMeanWashShareTxThousandOneToTenThousand}{21.65\% }
\newcommand{\NRandomMeanWashShareTxThousandOneToTenThousand}{2,295}

\newcommand{\PctRandomMeanWashShareTxTenThousandOneToHundredThousand}{50.31\% }
\newcommand{\NRandomMeanWashShareTxTenThousandOneToHundredThousand}{27}

\newcommand{\PctRandomMeanWashShareTxHundredThousandOnePlus}{0.00\% }
\newcommand{\NRandomMeanWashShareTxHundredThousandOnePlus}{0 }

\newcommand{\PctSelectedGradCohortZero}{1.13\% }
\newcommand{\NSelectedGradCohortZero}{139,614 }

\newcommand{\PctSelectedGradCohortOneToFive}{0.95\% }
\newcommand{\NSelectedGradCohortOneToFive}{1,365 }

\newcommand{\PctSelectedGradCohortSixToTwentyFive}{1.36\% }
\newcommand{\NSelectedGradCohortSixToTwentyFive}{2,282 }

\newcommand{\PctSelectedGradCohortTwentySixToHundred}{2.12\% }
\newcommand{\NSelectedGradCohortTwentySixToHundred}{2,217 }

\newcommand{\PctSelectedGradCohortHundredOneToFiveHundred}{2.98\% }
\newcommand{\NSelectedGradCohortHundredOneToFiveHundred}{2,281 }

\newcommand{\PctSelectedGradCohortFiveHundredPlus}{10.77\% }
\newcommand{\NSelectedGradCohortFiveHundredPlus}{882 }

\newcommand{\PctSelectedMeanWashShareTxOneToTen}{0.15\% }
\newcommand{\NSelectedMeanWashShareTxOneToTen}{63,512 }

\newcommand{\PctSelectedMeanWashShareTxElevenToHundred}{1.95\% }
\newcommand{\NSelectedMeanWashShareTxElevenToHundred}{67,651 }

\newcommand{\PctSelectedMeanWashShareTxHundredOneToThousand}{13.21\% }
\newcommand{\NSelectedMeanWashShareTxHundredOneToThousand}{15,660 }

\newcommand{\PctSelectedMeanWashShareTxThousandOneToTenThousand}{22.59\% }
\newcommand{\NSelectedMeanWashShareTxThousandOneToTenThousand}{1,799 }

\newcommand{\PctSelectedMeanWashShareTxTenThousandOneToHundredThousand}{34.72\% }
\newcommand{\NSelectedMeanWashShareTxTenThousandOneToHundredThousand}{19 }

\newcommand{\PctSelectedMeanWashShareTxHundredThousandOnePlus}{0.00\% }
\newcommand{\NSelectedMeanWashShareTxHundredThousandOnePlus}{0 }

\section{Wash trading}
\label{wash_trading_section}
In this section, we describe two heuristics for wash trading detection: one provides a conservative lower bound, while the other is more inclusive. 
We show that even under the conservative heuristic, a considerable share of coins exhibit wash-trading activity. 
Moreover, we find that, on average, for coins with more than 10,000 transactions, wash trading accounts for the majority of their transactions.  
Last, we evidence a strong correlation between wash trading activities and coin graduation. 

\subsection{Method}
We define two heuristics for wash trading detection, denoted as \textbf{WT1} and \textbf{WT2}. 
WT1 provides a conservative lower bound on wash trading prevalence, while WT2 captures a broader set of transactions that may be consistent with wash trading.

\begin{figure}[t]
    \centering
    \includegraphics[width=0.7\linewidth]{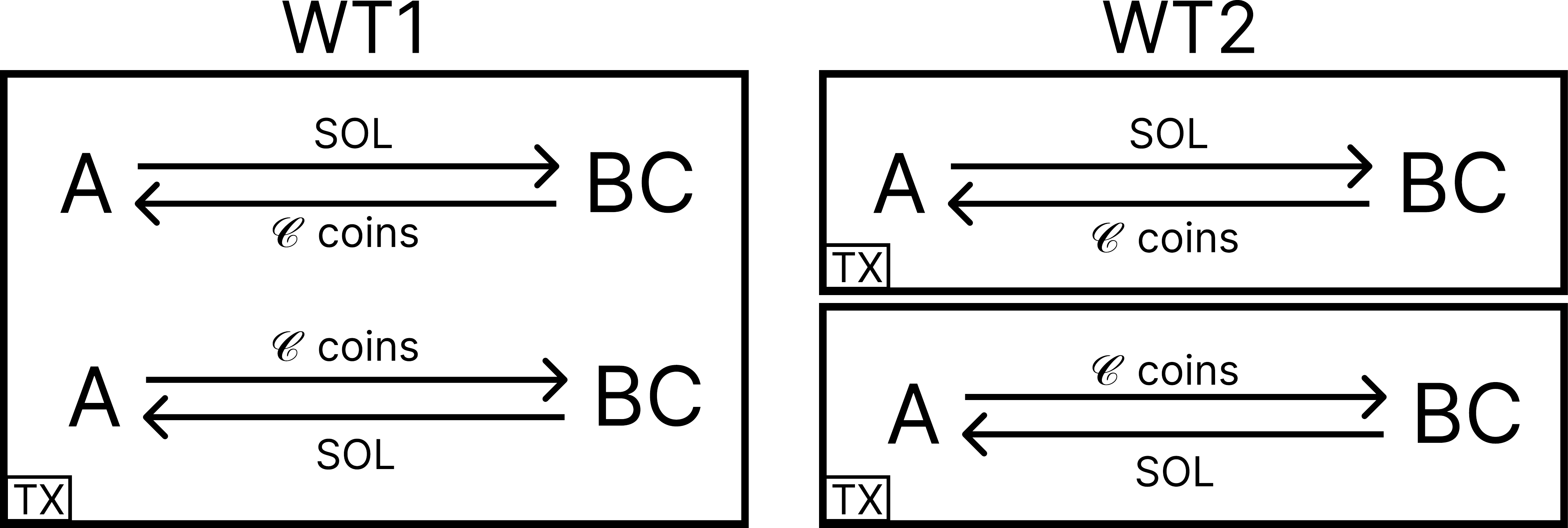}
    \caption{Wash trading heuristics representation.}
    \label{fig:wash_trading_heuristics_representation}

\end{figure}

\noindent \textbf{WT1}: 
As shown in Figure~\ref{fig:wash_trading_heuristics_representation}, we flag transactions in which address $A$ buys and sells the same amount $C$ \textbf{within a single transaction} against the same bonding curve $BC$ (i.e., the same coin). 
This heuristic captures transactions that meet the definition of wash trading (\S\ref{subsec:bg_manipulation_strategies}), namely those in which the same entity buys and sells atomically. 
We believe WT1 introduces an extremely small number of false positives since WT1 does not label MEV transactions~\cite{daian2020flash,qin2022quantifying} as wash trades; arbitrage or trading for another token is not possible within \pumpfun's bonding curve. 
Notably, WT1 only captures transactions executed outside the \pumpfun user interface (e.g., via custom smart contracts), since 
the \pumpfun website does not allow users to buy and sell within the same transaction, reflecting the manipulators' technical sophistication.

\noindent \textbf{WT2}: 
As shown in Figure~\ref{fig:wash_trading_heuristics_representation}, we flag transactions in which the same address buys and sells the same coin amount at \textit{approximately} the same price within a finite time window (whose length we will vary), 
regardless of whether these trades happen within a single transaction. 
As detailed in Appendix~\ref{appendix_pump_fun_trading_fees}, ``approximate'' accounts for the trading fees.
This heuristic relaxes the atomicity requirement of WT1. It captures the case where manipulators split the buy and sell trades across multiple consecutive transactions, with no other trade in between.
However, it can also include false positives, such as when an address buys a coin with the intent to profit, but (almost) no one else buys, so the address sells the coin back.

\subsection{Results}
\begin{figure*}[t]
    \centering

    \begin{subfigure}[t]{0.48\textwidth}
        \centering
        \includegraphics[width=\linewidth,clip]{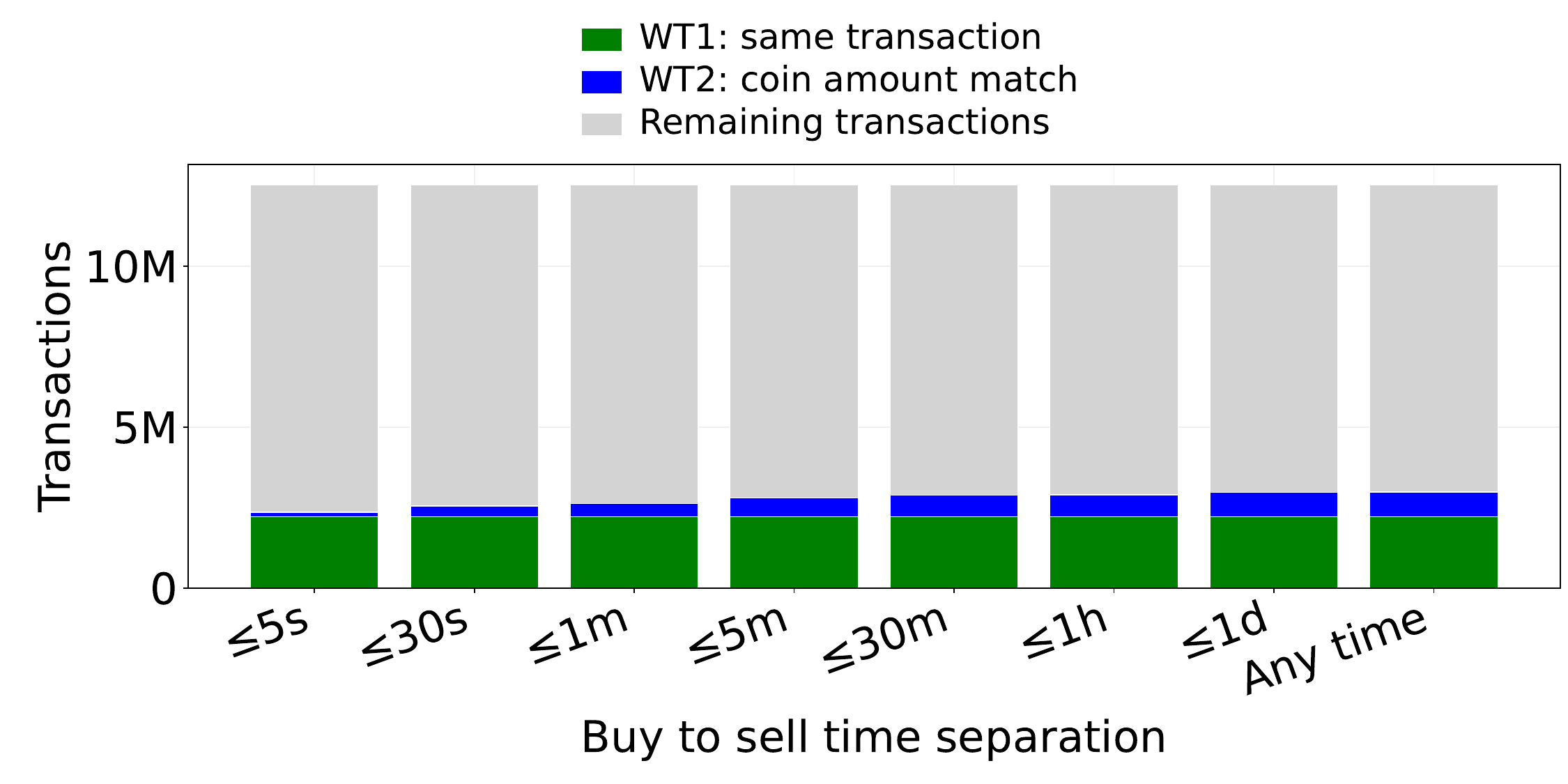}
\end{subfigure}\hfill
    \begin{subfigure}[t]{0.48\textwidth}
        \centering
        \includegraphics[width=\linewidth]{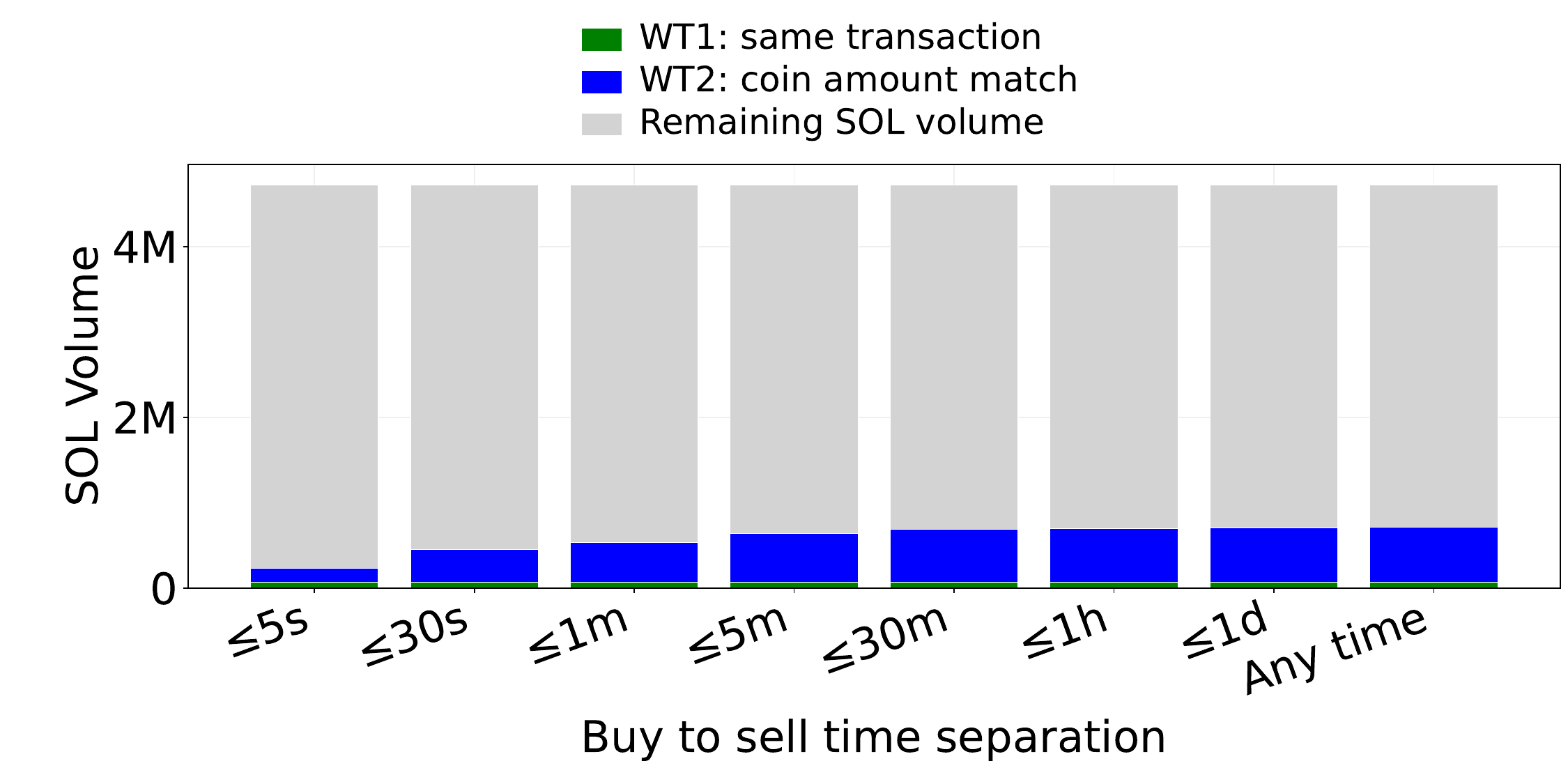}
\end{subfigure}

    \caption{Wash trading transaction count (left) and volume in SOL (right) under WT1 and WT2, across time thresholds (the $x$-axis), for the \onepctsample.}
    \label{tab:wash_trading_detection_results}

\end{figure*}

Figure~\ref{tab:wash_trading_detection_results} represents the number of wash-trading transactions (left) and the wash trading volume (right) for the \onepctsample. 
The $x$-axis represents the time lag between a manipulator's buy and sell. 
The green bars show transactions flagged by WT1 (i.e., no change across the $x$-axis), while the blue bars show those flagged by WT2 but not by WT1. 
We present only the results for the \onepctsample, as they are similar to those for the \fivedaysample.
We report the full results for the \fivedaysample in Appendix~\ref{appendix-5-day-sample-wash-trading}.
We see that increasing the buy-to-sell separation window (the $x$-axis) does not significantly capture more transactions. 

Under WT1, we find 2,221,734 wash-trading transactions across \PctRandomHOneWashTokens~of coins (i.e., coins with at least one wash-trading transaction). 
Under WT2 and a buy-to-sell time window of 1 second, we find 18,730 coins with wash trading instances, i.e., 12.38\% of coins.
Under WT2 and a 5-second time window, we find 2,356,189 wash-trading instances, corresponding to 44.05\% of coins. 
If we further increase the window size, the percentage of coins flagged by WT2 increases dramatically, reaching 88.5\% with a 1-day window. 
With WT2, while the number of wash trading candidate transactions does not grow significantly with the buy-to-sell time window, the number of affected coins, on the other hand, ramps up significantly, suggesting the presence of false positives.

We also quantify wash trading volume (in SOL).
WT1 accounts for \RandomHOneWashVolumeSol~SOL (\PctRandomHOneWashVolume of total trading volume and a median of \PctRandomHOneMedianWashShare for each coin's volume).  
The wash trading volume is significantly smaller than the wash trading count.
Manipulators appear to prioritize transaction count over volume to reduce the transaction fees since \pumpfun charges fees based on the trading volume.

We next analyze wash-traded coins.
We present two major results: 1) the proportion of wash-trading transactions is strongly correlated with the total number of transactions involving a given coin, and 2) the number of wash-trading transactions is strongly correlated with higher graduation rates.
For the rest of this analysis, we use the (conservative) results from WT1.

\noindent \textbf{Wash trading ratio per coin.}
In Table~\ref{tab:share_of_coins_flagged_by_wt1}, we report the ``wash-trading ratio,'' the share of wash-trading transactions among all transactions.
While the majority of wash-traded coins feature 100~or fewer transactions, we find that as the number of transactions for a given coin increases, the proportion of wash-trading transactions also increases strongly.
For instance, the wash-trading ratio is \PctRandomMeanWashShareTxThousandOneToTenThousand for coins with 1,000 to 10,000 transactions ($n=$\NRandomMeanWashShareTxThousandOneToTenThousand) and \PctRandomMeanWashShareTxTenThousandOneToHundredThousand for coins with more than 10,000 transactions ($n=$\NRandomMeanWashShareTxTenThousandOneToHundredThousand).

\begin{table}
\centering
    \caption{Nr. of total transactions and wash-trading ratio.}
    \begin{tabular*}{\columnwidth}{@{\extracolsep{\fill}}lrr@{}}
    \toprule
    No. of txs & Wash trading ratio (\%) & Nr. of coins \\
    \midrule
    1--10         & \PctRandomMeanWashShareTxOneToTen & \NRandomMeanWashShareTxOneToTen \\
    11--100       & \PctRandomMeanWashShareTxElevenToHundred & \NRandomMeanWashShareTxElevenToHundred \\
    101--1,000    & \PctRandomMeanWashShareTxHundredOneToThousand & \NRandomMeanWashShareTxHundredOneToThousand \\
    1,001--10,000 & \PctRandomMeanWashShareTxThousandOneToTenThousand & \NRandomMeanWashShareTxThousandOneToTenThousand \\
    10,001+       & \PctRandomMeanWashShareTxTenThousandOneToHundredThousand & \NRandomMeanWashShareTxTenThousandOneToHundredThousand \\
    \bottomrule
    \end{tabular*}
    \label{tab:share_of_coins_flagged_by_wt1}
\end{table}

\noindent \textbf{Graduation outcomes.}
We investigate how wash-trading transactions affect coin graduation.
We hypothesize that coins with many wash-trading transactions are more likely to graduate because they may appear more often on the top page of the \pumpfun web interface~\cite{luo2026resisting, kawai2024stranger}.
The graduation rate for the wash-traded coins (by WT1) is 2.0\% while that for the non-wash traded coins is 0.90\%. 
Table~\ref{graduation_by_number_of_transactions_matching_wt1} further illustrates that, when the number of wash-trading transactions is large, the graduation rates get higher.
To statistically examine the relationship between these two variables, we perform a logistic regression.
For coins with substantial wash-trading activity, when the number of wash-trading transactions doubles, the graduation odds increase by around 19\%. 
This relation is statistically significant ($p$-value: $3 \times 10^{-59}$).

\begin{table}[t]
\centering
    \caption{Graduation rates by the number of wash-trading transactions.}
    \begin{tabular*}{\columnwidth}{@{\extracolsep{\fill}}lrr@{}}
        \toprule
        Nr. of wash trading txs & Grad rate (\%) & Nr. of coins \\
        \midrule
        0        & \PctRandomGradCohortZero & \NRandomGradCohortZero \\
        1--5     & \PctRandomGradCohortOneToFive & \NRandomGradCohortOneToFive \\
        6--25    & \PctRandomGradCohortSixToTwentyFive & \NRandomGradCohortSixToTwentyFive \\
        26--100  & \PctRandomGradCohortTwentySixToHundred & \NRandomGradCohortTwentySixToHundred \\
        101--500 & \PctRandomGradCohortHundredOneToFiveHundred & \NRandomGradCohortHundredOneToFiveHundred \\
        500+     & \PctRandomGradCohortFiveHundredPlus & \NRandomGradCohortFiveHundredPlus \\
        \bottomrule
    \end{tabular*}
    \label{graduation_by_number_of_transactions_matching_wt1}
\end{table}

\noindent \textbf{Advanced atomic wash trading.}
We looked for more advanced wash-trading transactions to identify potential false negatives in WT1. 
Following Victor and Weintraud~\cite{victor2021detecting}, we searched for transactions in which an address $A$ buys coins, sends them to an address $B$, and $B$ sells them.
We did not find such transactions.
While we do not capture all advanced forms of wash trading (i.e., buying coins, splitting them, and selling them separately), manipulators appear to wash trade often naively (WT1).

\newcommand{\NClusterableMinters}{3,462,252 }
\newcommand{\PctClusterableMinters}{59.37\% }

\newcommand{\NOneHopClusteredAddresses}{2,797,830 }
\newcommand{\PctOneHopClusteredAddresses}{48.02\% }

\newcommand{\NOneHopClusters}{322,620 }
\newcommand{\PctLinkedMints}{62.99\% } 

\newcommand{\NTwoHopClusteredAddresses}{3,101,568 }
\newcommand{\PctLinkedAddrsTwoHop}{53.23\% }

\newcommand{\NTwoHopClusters}{295,978 }
\newcommand{\PctLinkedMintsTwoHop}{67.54\% } 

\newcommand{\NThreeHopClusteredAddresses}{3,203,502 }
\newcommand{\PctLinkedAddrsThreeHop}{54.98\% }

\newcommand{\NThreeHopClusters}{290,319 }
\newcommand{\PctLinkedMintsThreeHop}{68.78\% } 

\newcommand{\PctTopOneUnclusteredOutput}{38.89\% }
\newcommand{\PctTopFiveUnclusteredOutput}{52.85\% }
\newcommand{\PctTopTenUnclusteredOutput}{60.26\% }

\newcommand{\PctTopOneClustersOutput}{52.99\% }
\newcommand{\PctTopFiveClustersOutput}{67.63\% }
\newcommand{\PctTopTenClustersOutput}{74.37\% }

\newcommand{\PctTopOneClustersOutputTwoHop}{57.47\% }
\newcommand{\PctTopFiveClustersOutputTwoHop}{70.80\% }
\newcommand{\PctTopTenClustersOutputTwoHop}{76.90\% }

\newcommand{\PctTopOneClustersOutputThreeHop}{58.57\% }
\newcommand{\PctTopFiveClustersOutputThreeHop}{71.66\% }
\newcommand{\PctTopTenClustersOutputThreeHop}{77.67\%}

\newcommand{\MedianClusterSizeThreeHop}{3 } 
\newcommand{\MaxClusterSizeThreeHop}{45,833}

\section{Creator Address Obfuscation}
\label{sec:evaluation_clustering}
This section investigates creator address concentration and obfuscation.

\subsection{Method}
\label{address_clustering_description}
We cluster coin creator addresses using a ``common funder'' heuristic, following prior work~\cite{meiklejohn2013fistful, victor2020address}. 
Intuitively, addresses controlled by the same entity are often funded by a shared upstream address. 

We form a directed funding graph in which each coin-creating address $A$ has a directed edge from its funder $F(A)$---the address that sends the first SOL or token transfer to $A$.
Our approach yields a graph where each node has at most one incoming edge. 
To avoid spurious aggregation, we remove any node and its associated edges from this graph if Arkham's API labels the node as a third-party service (e.g., exchanges, bridges, smart contracts).

We then construct clusters by traversing the funding graph starting from coin-creator addresses up. 
For each such address $A$, we recursively follow the funder relation $F(A)$ to identify a chain of funders.
We denote a cluster linking coin creators $A$ to their funders $F(A)$ a 1-hop cluster. 
Similarly, we define 2-hop clusters as clusters linking coin creators to their funders' funders, i.e., $F(F(A))$, and 3-hop clusters by $F(F(F(A)))$. 
We stop at 3-hop as we observe minimal coverage increase when going from 2-hop to 3-hop, as documented in Appendix~\ref{appendix_creator_address_obfuscation}.

While Arkham claims that their verified labels have 98\% accuracy~\cite{arkham_labeling_guide}, our approach erroneously links two unrelated addresses if Arkham misses labeling an address belonging to third-party services. 
This problem tends to exacerbate as we increase the funding hops~\cite{victor2020address}.
We therefore check whether the existence of a single address that would dominantly fund the rest of the addresses skews our results.
We perform an ``adversarial leave-one-address-out'' analysis, where we remove each address in the cluster one by one and check how much of the cluster's connectivity remains after the removal. 
We then identify the address that breaks the cluster the most (i.e., the worst-case deletion).  
We use \emph{cluster connectivity retained} to assess the cluster connectivity after the worst-case deletion.
It is defined as the share of addresses that belong to the largest component (i.e., the connected part).

\subsection{Results}
We apply the approach described in \S\ref{address_clustering_description} to cluster \NUniqueCreators unique coin creator addresses.

\noindent \textbf{Address concentration and obfuscation.}
Out of \NUniqueCreators unique coin creator addresses, over half of them are assigned to a multi-address cluster (i.e., we manage to find the common funder). 
Address-level counts may substantially overestimate the number of distinct coin creators. 
We find \NOneHopClusters multi-address clusters for 1-hop clustering, \NTwoHopClusters for 2-hop clustering, and \NThreeHopClusters for 3-hop clustering. 
Among addresses assigned to multi-address clusters (i.e., 3,203,502 addresses), our clustering (3-hop) reduces the apparent number of distinct creators by roughly 11 times.

Even under the conservative 1-hop clustering setting, multi-address clusters are nontrivial in size: the median multi-address cluster contains 3 addresses, and the largest cluster contains 10,531 addresses. 
These results indicate that a substantial fraction of coin creation activity is spread across multiple addresses, suggesting frequent address reuse and possible deliberate obfuscation.

\noindent \textbf{Coin creation concentration.}
We observe that multi-address clusters account for a disproportionate share of coin creation. 
Under 1-hop clustering, addresses in multi-address clusters account for \PctLinkedMints of all coins, even though they represent only \PctOneHopClusteredAddresses of all coin creator addresses.

\begin{figure}[t]
\centering
\includegraphics[width=0.9\columnwidth]{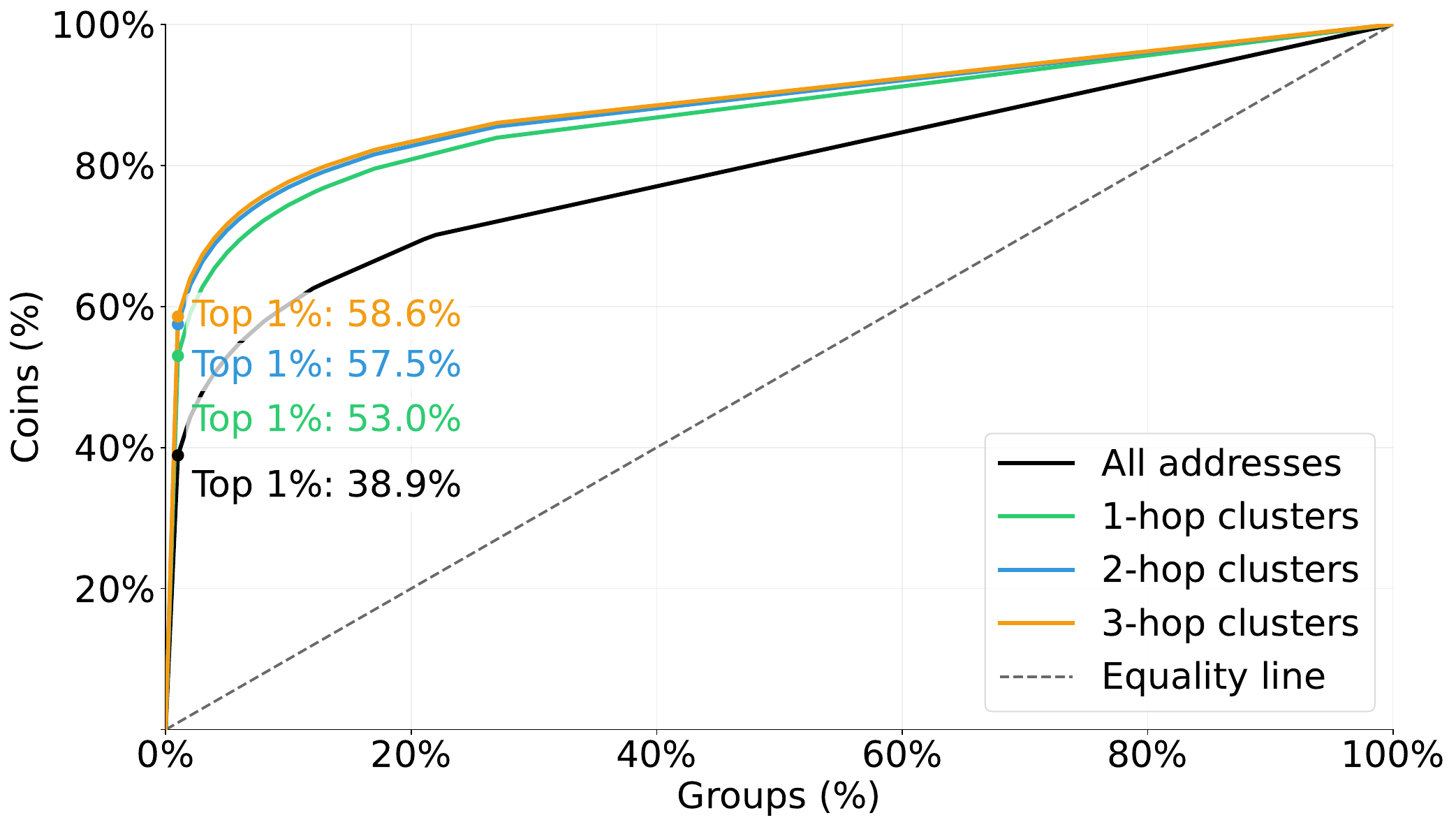}
\caption{Fractions of coins (the $y$-axis) created by the top $x$\% of clusters (including singletons).}
\label{fig:cluster_pareto}
\end{figure}

We note similar patterns when considering the overall concentration of coin creation. Figure~\ref{fig:cluster_pareto} shows the fraction of coins created by the most active clusters. Without clustering, the top 1\% of coin creator addresses account for \PctTopOneUnclusteredOutput of all coins. After clustering, this concentration increases substantially. Under 1-hop clustering, the top 1\% of clusters account for \PctTopOneClustersOutput of all coins. The concentration further increases under 2-hop and 3-hop clustering: the top 1\% of clusters account for \PctTopOneClustersOutputTwoHop and \PctTopOneClustersOutputThreeHop of all coins, respectively.

Some of the largest clusters are highly active from late Feb. to May 2024 onwards, while others exist for only a certain period of time (e.g., 240 days for the 11th-largest cluster, just 9 days for the 38th-largest cluster).
On average, the top 100 clusters (i.e., clusters with at least 2,464 creators) are active for 434 days. 

\noindent \textbf{Cluster validation.}
We perform validation analyses to assess whether the observed clusters genuinely reflect multi-address activity (e.g., rather than spurious links introduced by unlabeled third-party services).

First, we examine the distribution of coin creation within the top 1\% most active clusters, ranked by the number of coins created. 
In the median cluster, the most active address accounts for 26.33\% of the cluster's coins, while the three most active addresses together account for 50.33\%. 
Conversely, the median number of addresses needed to account for 90\% of a cluster's coins is 25.5. 
These results suggest that a single dominant address does not typically explain high-output clusters.
Instead, coin creation is distributed across many addresses within the cluster.

Second, we evaluate cluster connectivity through adversarial deletion defined in \S\ref{address_clustering_description}. 
For the top 1\% most active clusters, after the worst-case deletion, the largest remaining component retains a median of 80.4\% of the cluster's addresses and 89.4\% of its coin output. 
Our validation results indicate that the concentration patterns documented above are unlikely to be driven solely by a single address (e.g., unlabeled third-party entity).

\newcommand{\PctAllMultiWalletLargestShare}{50.00\% }
\newcommand{\PctAllMultiWalletTopThreeShare}{100.0\% }
\newcommand{\NAllMultiWalletsForNinety}{2.0 }

\newcommand{\PctTopTenOutputLargestShare}{37.74\% }
\newcommand{\PctTopTenOutputTopThreeShare}{68.70\% }
\newcommand{\NTopTenOutputWalletsForNinety}{8.0 }

\newcommand{\PctTopFiveOutputLargestShare}{32.83\% }
\newcommand{\PctTopFiveOutputTopThreeShare}{60.91\% }
\newcommand{\NTopFiveOutputWalletsForNinety}{12.0 }

\newcommand{\PctTopOneOutputLargestShare}{26.33\% }
\newcommand{\PctTopOneOutputTopThreeShare}{50.33\% }
\newcommand{\NTopOneOutputWalletsForNinety}{25.5 }

 \newcommand{\NSelectedAtomicDumps}{3,510 }
\newcommand{\NSelectedAtomicDumpTokens}{1,114 }
\newcommand{\PctSelectedAtomicDumpTokens}{0.75\%}

\newcommand{\NSelectedWindowDumps}{9,793 }
\newcommand{\NSelectedWindowDumpTokens}{4,451 }
\newcommand{\PctSelectedWindowDumpTokens}{2.99\% }

\newcommand{\NRandomAtomicDumps}{4,402 }
\newcommand{\NRandomAtomicDumpTokens}{1,239 }
\newcommand{\PctRandomAtomicDumpTokens}{0.81\%}

\newcommand{\NRandomWindowDumps}{11,005 }
\newcommand{\NRandomWindowDumpTokens}{5,080 }
\newcommand{\PctRandomWindowDumpTokens}{3.34\%}

\newcommand{\SelectedAtomicDumpVolumeSol}{29,608 } \newcommand{\SelectedAtomicDumpProfitSol}{-1,856 } 

\newcommand{\RandomAtomicDumpVolumeSol}{33,393 } \newcommand{\RandomAtomicDumpProfitSol}{-2,197 } 

\newcommand{\SelectedWindowDumpVolumeSol}{42,995 } \newcommand{\SelectedWindowDumpProfitSol}{-1,393 } 

\newcommand{\RandomWindowDumpVolumeSol}{50,196 } \newcommand{\RandomWindowDumpProfitSol}{-2,109} 

\newcommand{\PctSelectedAtomicDumpVolume}{0.30\% }
\newcommand{\PctSelectedWindowDumpVolume}{0.43\% }

\newcommand{\PctRandomAtomicDumpVolume}{0.33\% }
\newcommand{\PctRandomWindowDumpVolume}{0.50\% }

\newcommand{\MedianAtomicSourcesPerDump}{7 }
\newcommand{\MaxAtomicSourcesPerDump}{7}

\newcommand{\MedianWindowSourcesPerDump}{7 }
\newcommand{\MaxWindowSourcesPerDump}{17}

\section{Coordinated sell}
\label{sec:evaluation_dumping}
This section investigates the scenario where a manipulator 1) deploys multiple addresses that buy tokens individually, 2) funnels all bought tokens to a single address, and 3) sells all tokens at once to profit. 
In particular, we focus on the ``dumping'' part (2, 3) that often happens consecutively.

\subsection{Method}
We define two heuristics: \textbf{DP1} and \textbf{DP2}, to identify coordinated sell, or ``dumps.'' 
DP1 provides a conservative lower bound, while DP2 captures a broader set of patterns. 

\begin{figure}[t]
    \centering
    \includegraphics[width=0.7\linewidth]{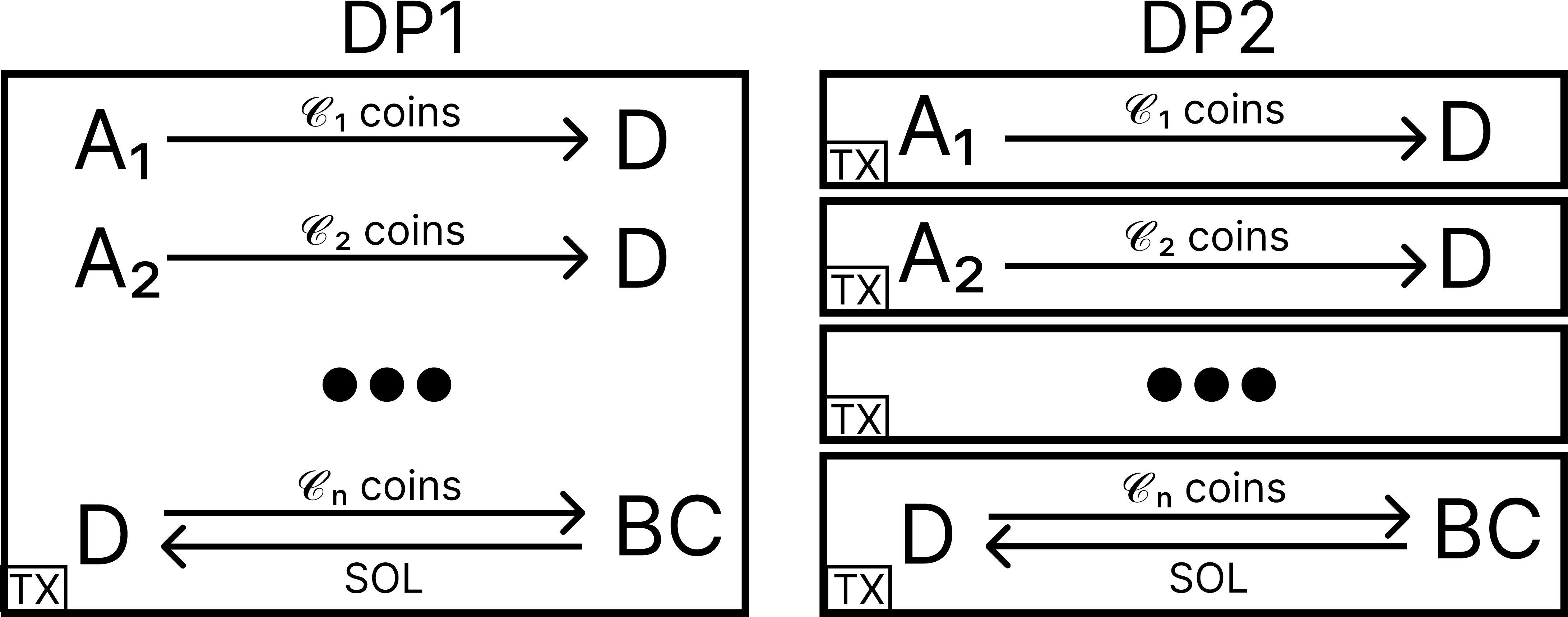}
    \caption{Dumping heuristics representation.}
    \label{dumping_heuristics_representation}
\end{figure}

\textbf{DP1}:
In Figure~\ref{dumping_heuristics_representation} (left), DP1 captures the most direct form of coordinated dumping.
Multiple ``sender'' addresses $A_1, A_2, ...$ transfer tokens to a single ``dumper'' address $D$, which sells all received coins.
All of this happens atomically, i.e., \textbf{within a single transaction}.
Transactions that meet these criteria are unlikely to yield false positives. 
Indeed, all senders need to sign the transaction with their private keys---a strong sign that they belong to the same entity (or at the very least, colluding entities).
Furthermore, manipulators need to directly interact with the blockchain to execute this, just like \textbf{WT1} in \S\ref{wash_trading_section}.

\textbf{DP2}: 
In Figure~\ref{dumping_heuristics_representation} (right), we relax the atomicity requirement, allowing for a time window between the transfers and the sell. 
We collect all coin transfers from $A_1, A_2, ...$ to $D$ that occur within 24 hours of $D$'s dumping.
If $D$ receives coins from at least two distinct senders and then sells at least the received amount, we classify it as a DP2 dump. 
When the window size is large, this heuristic may introduce false positives where two distinct traders happen to transfer their tokens to another trader, who sells them later. 

\begin{figure}[t]
    \centering
    \includegraphics[width=\linewidth]{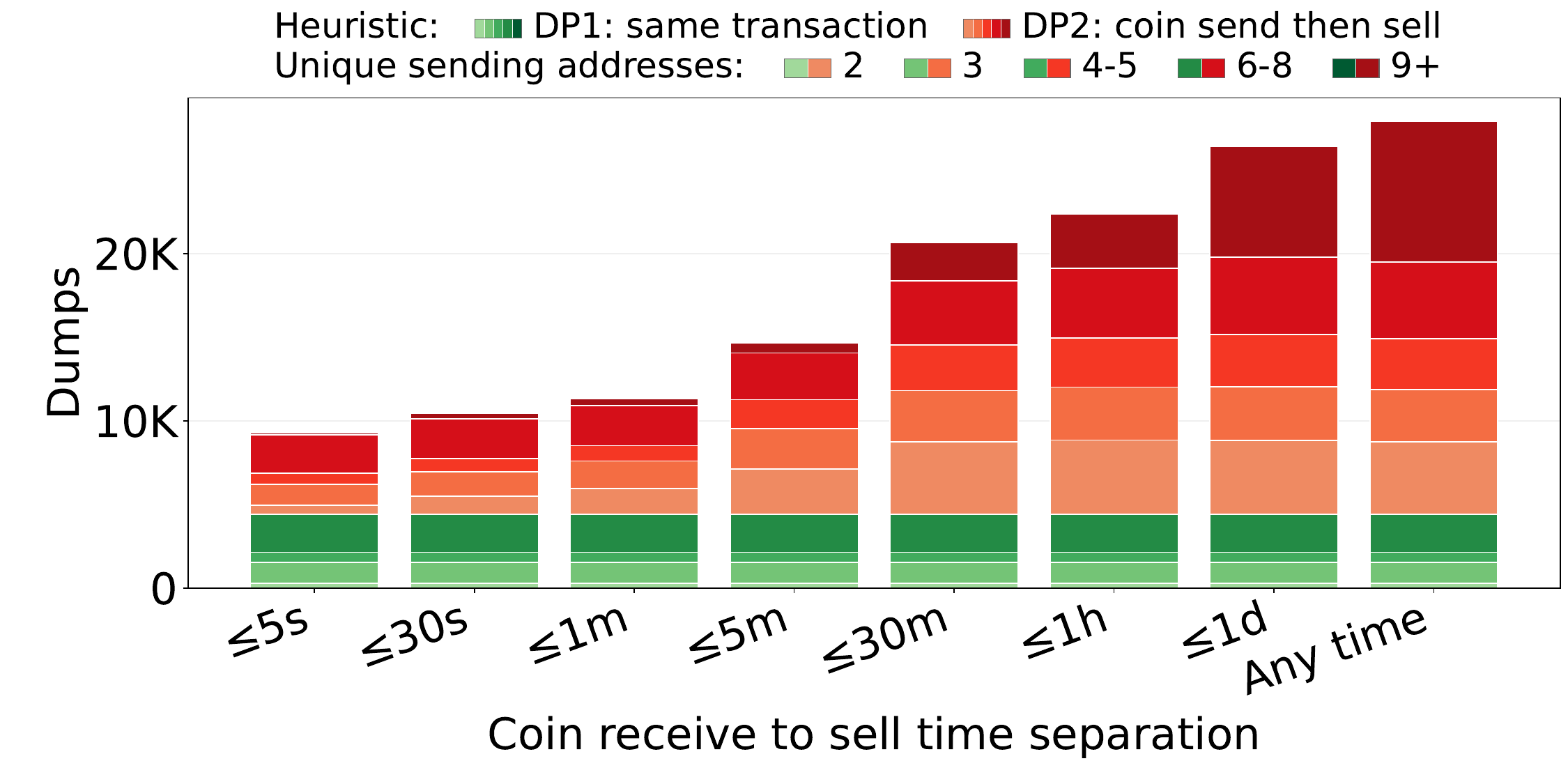}
    \caption{Nr. of dump instances for DP1 and DP2 across time thresholds (the $
    x$-axis) for \onepctsample}
    \label{fig:dump_evaluation_a}
\end{figure}

\subsection{Results}
Just like in \S\ref{wash_trading_section}, we only report our findings for the \onepctsample. 
Results for the \fivedaysample are similar, and can be found in Appendix~\ref{appendix-5-day-sample-dumping}.
Figure~\ref{fig:dump_evaluation_a} shows the number of dumping instances (i.e., transfers and a sell pair) across different time window sizes (the $x$-axis).
The green bars indicate the dumping instances flagged by DP1 (i.e., no change across the $x$-axis), and the red bars are the dumping instances flagged by DP2 (but not by DP1). 
The color gradation corresponds to the number of senders involved, as shown in the legend.

We find that dumping is prevalent.
Under DP1, we detect \NRandomAtomicDumps dumps.
Under DP2, with time windows of 5 seconds, we capture 9,270 dumps.
The number increases to 66,706 and 88,730 dumps for 1 hour and 1 day, respectively. 
We also show that there are many senders involved in dumping.
Under DP1, the median number of senders is~\MedianAtomicSourcesPerDump (max~\MaxAtomicSourcesPerDump). 
Under DP2, the median number is \MedianWindowSourcesPerDump (max \MaxWindowSourcesPerDump).
In the \fivedaysample, we find one extreme case, a dump involving 312 senders, highlighting a high degree of coordination.

\begin{figure}
\centering
\includegraphics[width=0.7\columnwidth]{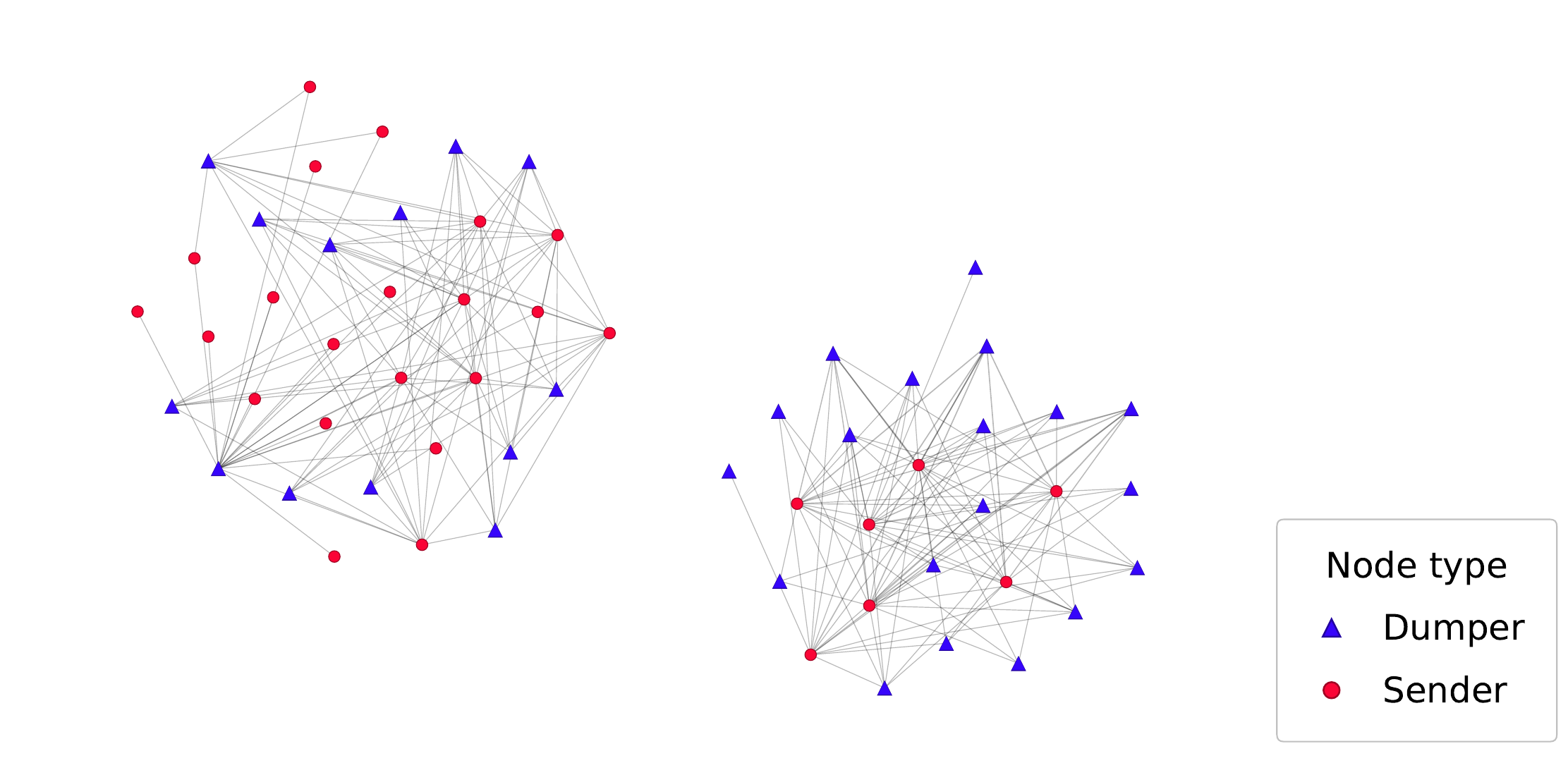}
\caption{Token transfers (edges) between senders and dumpers (nodes) in the two largest clusters.}
\label{fig:dump_evaluation_network}
\end{figure}

We next show that manipulators reuse senders and dumpers.
We construct a graph where each node is an address (either a sender or a dumper) and each edge represents a transfer from a sender to a dumper. 
Figure~\ref{fig:dump_evaluation_network} shows the graphs corresponding to the largest two clusters flagged by DP1.
Many senders repeatedly interact with many dumpers; the manipulators employ the same senders and dumpers repeatedly across manipulations.

Finally, we investigate dumping transaction volume under a conservative lower bound. Using DP1, the dumpers sell 33,393 SOL in total. In addition, considering the range of SOL price, which was roughly averaging USD~160 and never went below USD~80, we can estimate that the total dump volume is between USD~2.5M and~5M in this 1\% sample.

 \section{Copycat coins}
\label{sec:copycat}
We describe how we detect ``copycat'' coins (i.e., coins impersonating another asset), and the strategies surrounding them. 

\subsection{Methods}
\label{subsec:copycat_method}

We use three heuristics for identifying the set of original and copycat coins. 
The first and strictest approach detects coins with a shared \textbf{name}, \textbf{symbol}, \textbf{description} (can be empty), and \textbf{profile image}.
The second heuristic detects coins with a shared \textbf{name}, \textbf{symbol}, and \textbf{profile image}, because the description is not mandatory and is the easiest to modify. 
The third and broadest approach detects coins that share only \textbf{name} and \textbf{symbol}, since a manipulator can also crop or edit an image or change a file extension. 
To identify image duplicates, we rely on the structure of IPFS, which includes the file hash in the URL\footnote{IPFS generates a different identifier even if an image differs by a single pixel.} \cite{ipfs2026cid}. 
This approach gives us a straightforward mechanism for capturing exact image duplicates and mitigates risks (e.g., downloading offensive or illegal materials), but limits our ability to detect image similarity.

We define the original coin as the coin with the earliest creation timestamp and label the rest as copycats.
If the two coins share an identical creation timestamp, we consider the coin with the higher current market cap to be the original coin.
Finally, we exclude any pair of original and copycat coins that are created by the same address or belong to the same creator cluster (derived in \S\ref{sec:evaluation_clustering}), since such pairs likely belong to the same entity and do not exhibit impersonation intent.
We separately analyze these coins with identical metadata and creator in Appendix~\ref{appendix:clone_coins}.

\subsection{Results}
\label{subsec:copycat_result}

\begin{table}[]
\caption{Copycat detection results by the three heuristics}
\begin{adjustbox}{width=\linewidth,center}
\begin{tabular}{@{}lllll@{}}
\toprule
           & Nr. originals & Nr. copycats (ratio) & Ori. no suffix rate & Copy. no suffix rate \\ \midrule
1          & 647,986       & 1,490,123 (0.10)     & 0.14                & 0.26                 \\
2          & 757,930       & 1,866,178 (0.12)     & 0.14                & 0.25                 \\
3          & 1,270,892     & 5,392,081 (0.36)     & 0.14                & 0.20                 \\ \bottomrule
\end{tabular}
\end{adjustbox}
\label{tab:copycat_detection}
\end{table}

Table~\ref{tab:copycat_detection} summarizes the detection results for all three heuristics. 
The first and second methods capture over 1.5 million and 1.9 million copycats (10\%, 12\% of all coins), respectively. 
Most copycats that are caught by the second heuristic but not by the first either leave the description empty, or have a description associated with a token-deploying platform, e.g., ``Deployed using https://[Service-C] ($n=9,047$).

The broad third method identifies 5.4 million copycats (36\%) that share the same name and symbol.
This increase in detected coins comes from 1) copycat coins that modify either the description or image, or both, or 2) false positives of coins that happen to use the same name and symbol.
The second case often appears when coin creators use the same concept (e.g., coins created after celebrities).
We do not capture coins with slight modifications to their names and/or symbols, as seen in typosquatting research~\cite{moore2010measuring,kintis2017hiding,gabrilovich2002homograph}. 
As a conservative bound, we use the first heuristic for the rest of the analyses.

\begin{figure}
    \centering
    \begin{minipage}{0.48\linewidth}
        \centering
        \includegraphics[width=1\linewidth]{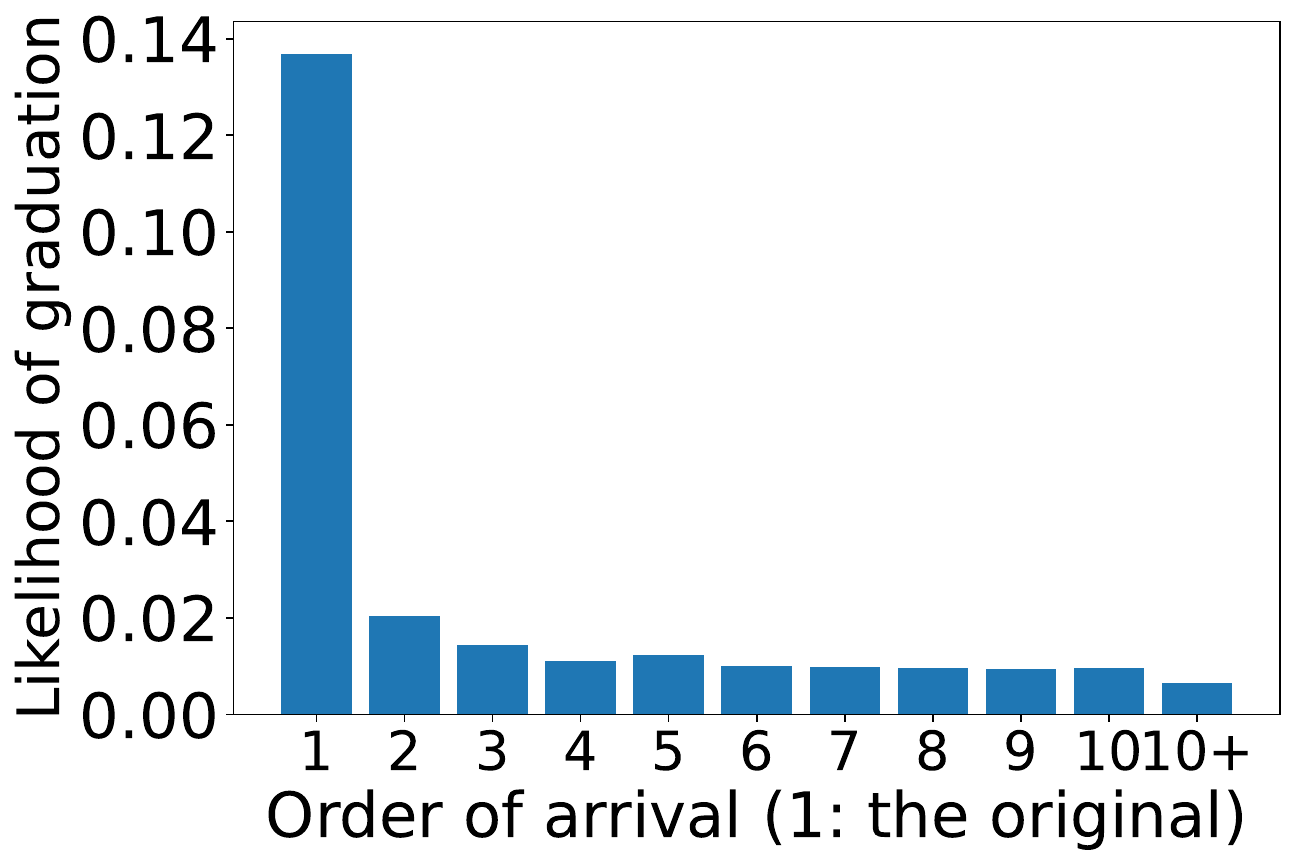}
    \end{minipage}
    \hfill
    \begin{minipage}{0.48\linewidth}
        \centering
        \includegraphics[width=1\linewidth]{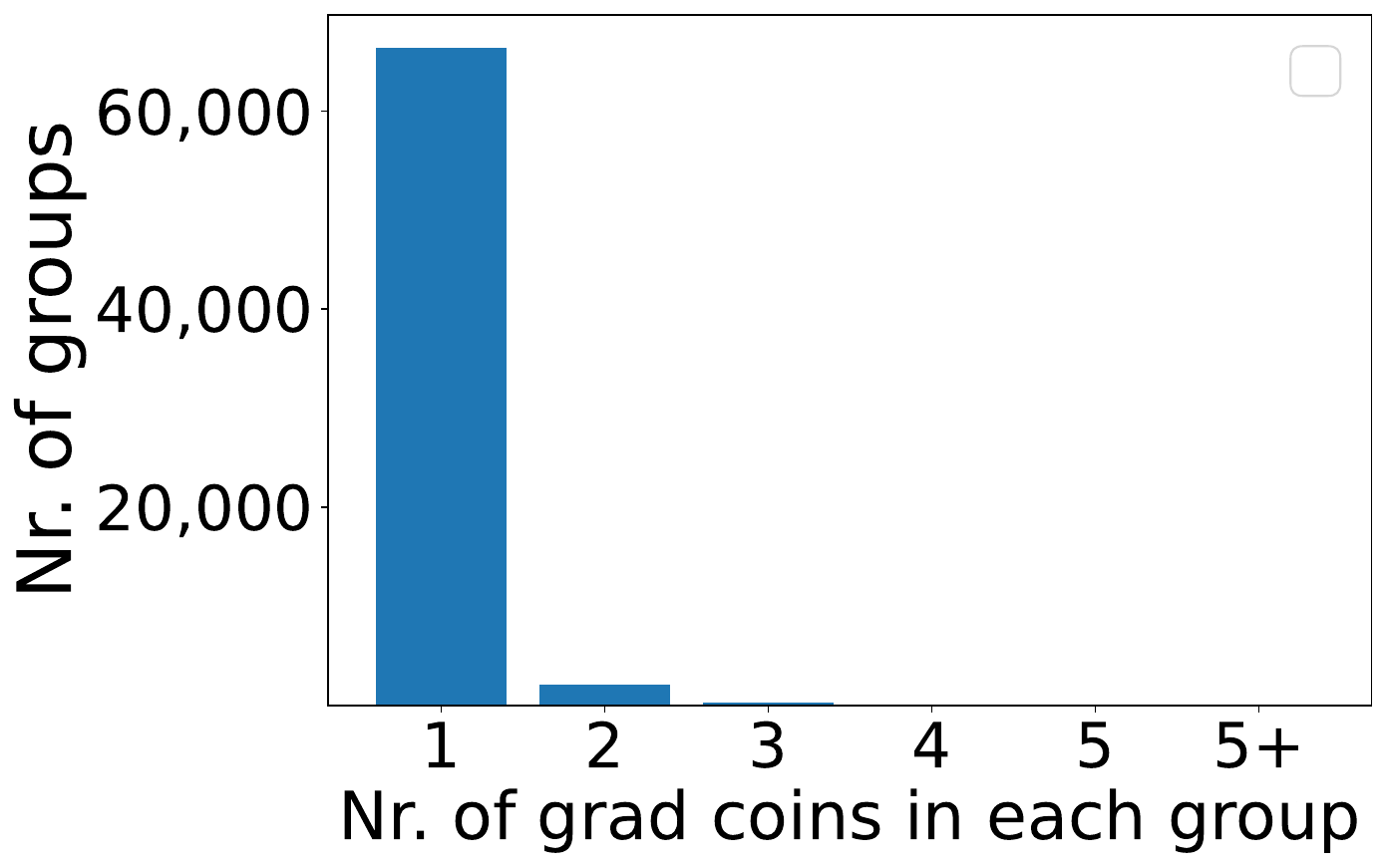}
    \end{minipage}
    \caption{1) Order of arrival and graduation likelihood and 2) Distributions of groups by nr. of graduated coins.}
    \label{fig:copycat_num_grad_latency}
\end{figure}

We first investigate the competition between originals and copycats.  
The original coins have a significantly higher graduation rate, 9.20\% (compared to the baseline 1.02\%), whereas the copycat coins are less likely to graduate, 0.86\%.
The signal here is that the presence of copycats is an endorsement of the 
original coin, reflected in the higher-than-average graduation rate of originals. 
In Figure~\ref{fig:copycat_num_grad_latency} (left), the $x$-axis is the order of arrival ($x=1$ is the original, $x=2$ is the earliest copycat), and the $y$-axis is the likelihood of graduation (\S\ref{sec:bg}). 
In this figure, we only include cases with at least 9 copycats. Becoming the original coin confers a significant first-mover advantage, and this benefit quickly disappears. 
Nevertheless, 17.7\% of graduated coins are copycats. Earlier copycats have a slightly higher probability of graduation than later ones.

We also examine if the original and the copycat(s) can both graduate.
Figure~\ref{fig:copycat_num_grad_latency} (right) further illustrates the number of graduated coins per ``group'' (i.e., the original and its copycats combined).
68,985 (10.65\%) of groups graduate at least one coin.
Only 2,613 groups manage to graduate multiple coins, accounting for just 0.40\% of the total groups.
While competition usually eliminates copycats, an original and the copycat(s) occasionally split users' attention and both graduate.

We will next estimate the level of automation present in the deployment of 
copycats.  
When coin creators launch a coin through the website's UI,
\pumpfun 
has been generating a vanity address~\cite{soska2021bitmex}, which ends with ``pump'' (lowercase) since (roughly) May 30, 2024.
Thus, if a coin address ends with a random string (``no suffix''), it means the creator directly interacted with a blockchain using their custom smart contract. 
Since the manipulators could brute-force an address ending with ``pump'' (requiring about $58^{4}~\approx11$ million attempts on average, which can be done in less than a second with the latest GPUs~\cite{tsuchiya2025address}), we consider our estimate as a lower bound.
Table~\ref{tab:copycat_detection} shows that 13.7\% of the original coins' addresses do not have a ``pump'' suffix. On the other hand, 26.5\% of the copycat coins' addresses do not include the ``pump'' suffix. This result indicates copycat deployment is much more automated than regular coin creation:  
for comparison, 
the baseline rate of addresses without the ``pump'' suffix, over all coins, is 16.05\%.

 \section{Social manipulation}
\label{sec:social_manipulation}
This section covers two channels that actors use for influence: 1) \pumpfun comments and 2) social media.

\subsection{Comment manipulation}
\label{subsec:comment_manipulation}
We first investigate automated bots that post promotional comments on \pumpfun's coin profile pages.

\noindent\textbf{Methods:}
Traders incorporate social signals (e.g., comments about coins) into their future price expectations.\footnote{A line of pump-and-dump studies confirms the impact of external channels (e.g., Telegram or Discord groups) on coin prices~\cite{mirtaheri2021identifying,tsuchiya2021profitability}.}
Strategic actors can therefore influence the outcome of a coin by using multiple accounts to ``astroturf'' publicly available social signals.

\begin{figure}
    \centering
    \begin{minipage}{0.4\linewidth}
        \centering
        \includegraphics[width=1\linewidth]{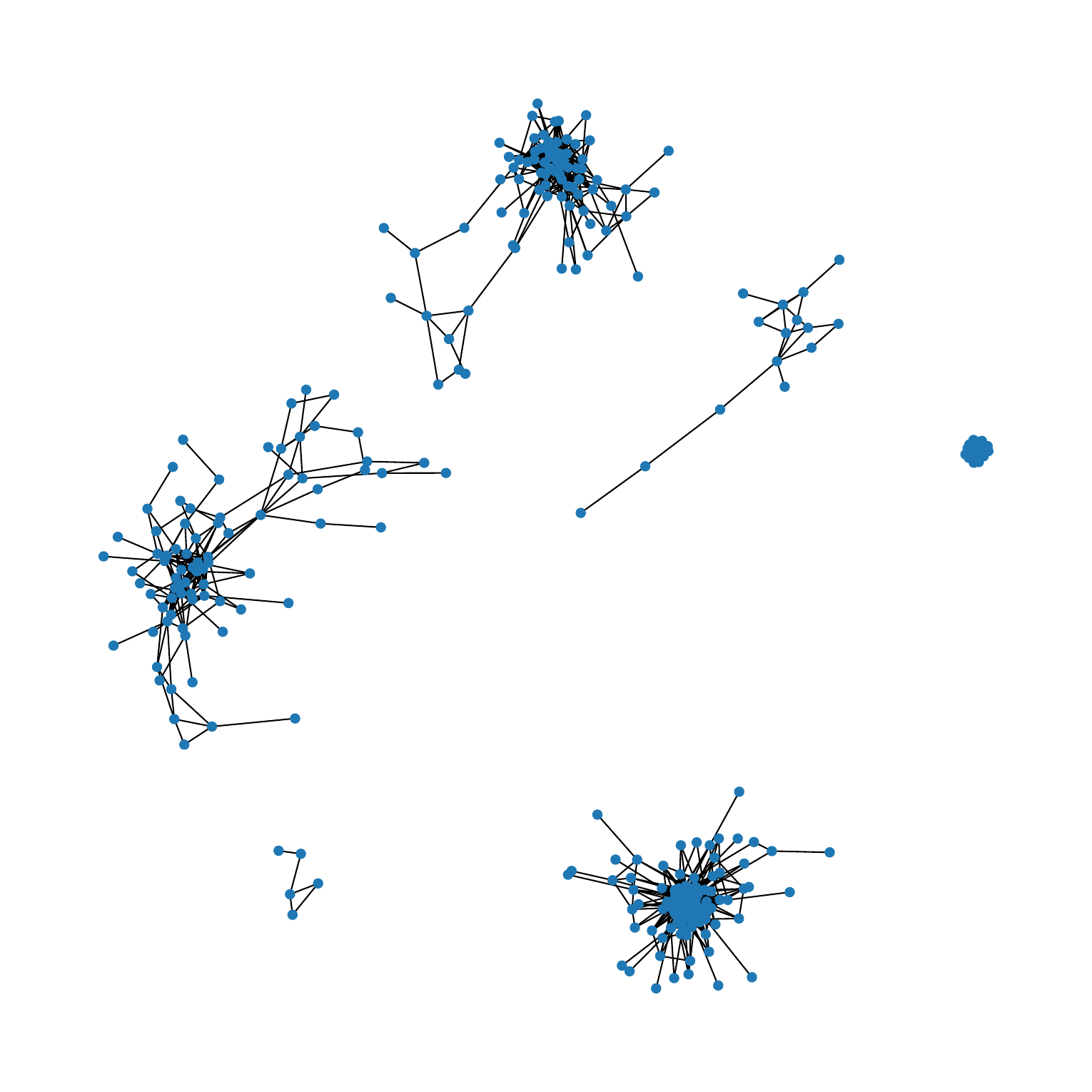}
    \end{minipage}
    \hfill
    \begin{minipage}{0.55\linewidth}
        \centering
        \includegraphics[width=1\linewidth]{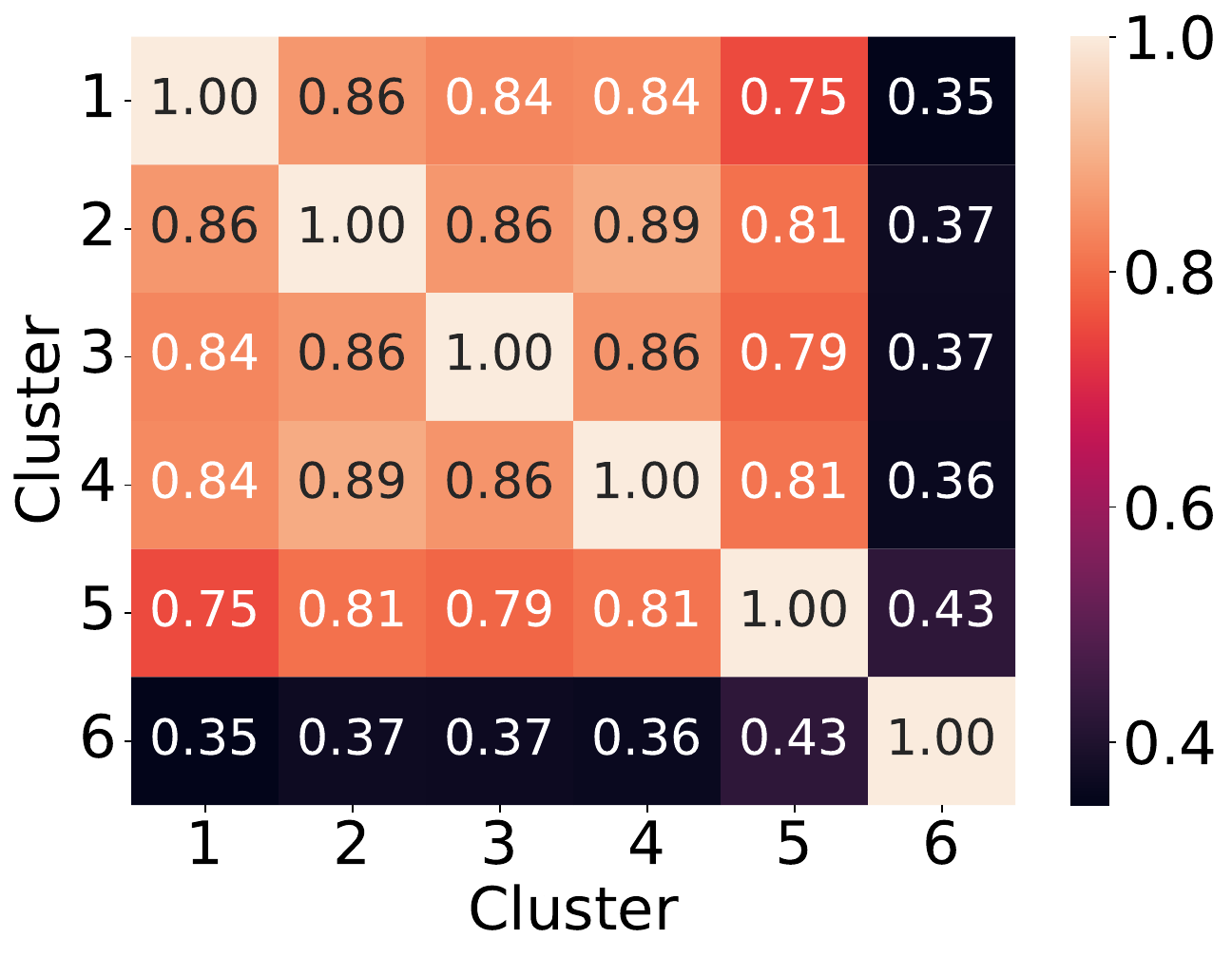}
    \end{minipage}
    \caption{(Left) user collusion visualization; (Right) text similarity among the clusters.}\label{fig:comment_network}
\end{figure}

\pumpfun provides a chatbox on each coin's profile page where users can freely post comments from their blockchain addresses; they do not need to own the token.
Using the comments from the 1\% coin sample, we construct a graph (Figure~\ref{fig:comment_network}, left) in which all nodes represent users (i.e., blockchain addresses), following the ``lockstep'' approach used in He et al.~\cite{he2026six}.
An edge between two users indicates that both users posted within $\tau=0.1$~seconds of each other on at least \emph{two} different coin pages. 
We also confirm that the graph remains structurally the same as $\tau$ varies from 0.1s to 1s. 
We exclude clusters with fewer than three nodes for visibility.

\noindent\textbf{Results:}
Figure~\ref{fig:comment_network} (left) shows six closely-knit clusters: 111, 80, 78, 25, 15, and 5~users, respectively, suggesting potential collusion to boost social engagement. 
We expect to find more linkages or a denser central cluster if we expand our dataset beyond the 1\% coin sample. 
While Tsuchiya et al.~\cite{tsuchiya2023misbehavior} find that reputation manipulation on financial articles uses \textit{short} positive comments, most messages from these clusters are longer (median: about 23 characters for most groups) and appear to mimic authentic user comments. 
The common practice includes capitalization for emphasis (e.g., ``HOW LONG I'VE WAITED FOR THIS! I BOUGHT IT AND I'M WAITING FOR THE PUMP.''), slang, abbreviations, typos (e.g., ``lmaoooo the idea is crazy , have u cjeck the website guys ?''). 
While most comments are positive, manipulators appear to intentionally include negative comments (e.g., ``That 10\% splitted wallet is not good,'' ``I smell a rug'').

We next examine comment similarity across clusters. 
Figure~\ref{fig:comment_network} (right) shows the Jaccard similarity, i.e., the fraction of text shared by all pairs of the six clusters. 
We see significant overlap in the text used \emph{across} clusters, especially among the top five clusters, which suggests two plausible explanations.
Each cluster could represent different actors that rely on the same software or services identified in \S\ref{sec:mmaas}.
The fact that they are both tight-knit and distinct from each other, and that this remains robust to timing threshold changes ($0.1 \leq \tau \leq 1$ seconds) support this interpretation.
Alternatively, all clusters could belong to the same entity: using 
a 1\% sample of all coins could have caused us to miss linkages.

These results suggest that content-based moderation (e.g., detecting positive short comments) advocated earlier~\cite{tsuchiya2023misbehavior} may no longer be effective.
Metadata-level detection (e.g., when and where users post) appears to be useful, especially given the rise of LLMs.

\subsection{Manipulations from social media}
\label{subsec:external_platforms}
We next investigate two kinds of manipulation strategies that influence social media signals about coins.

\noindent \emph{Community-based manipulation.} A manipulator can create a public or private channel on social media to advertise their coins or invite users to coordinate manipulation. 
We look at Telegram groups and Twitter communities whose links appear in \pumpfun coin metadata and analyze their strategies and outcomes.

\noindent \emph{Post-based manipulation.} Social media ``influencers'' (i.e., users with highly visible profiles) can post about something ``memeable,'' and simultaneously create a meme coin to profit from it. Even if a manipulator does not hold any social status, they can closely monitor influential social media posts and immediately launch related coins. 

For both manipulation types, 
our data cannot assert malicious \emph{intent} to coordinate price manipulation or extract money.
However, looking at market outcomes (e.g., graduation, price jumps) allows us to estimate how much a manipulator could \emph{potentially} reap.

\noindent\textbf{Methods:}
On \pumpfun, coin creators can write anything in three (optional) open fields: ``\texttt{twitter},'' ``\texttt{telegram},'' and ``\texttt{website}'' for each coin. 
Notably, \pumpfun does not validate the entered content (e.g., a Twitter username) against the corresponding services. 
We first use the \texttt{tldextract} library~\cite{tldextract} to extract domain names from the links creators include in each field.
59.6\%, 20.6\%, 43.8\% of coins link domains in the \texttt{twitter}, \texttt{telegram}, and \texttt{website} fields, respectively.
This result indicates the importance of studying external influences outside \pumpfun.  
We then aggregate all domain names (e.g., \texttt{youtube.com} and \texttt{youtu.be}) for each platform.
We report the aggregated statistics for linked external platforms in Appendix~\ref{appendix:external_platform_count}.

We next focus on Twitter and Telegram, accounting for the majority of coins  in the social media fields, and extract users, groups, and posts. 
We also look at Truth Social whose structure is highly similar to Twitter's.

Twitter links are either users, communities (i.e., public chat groups), or specific user posts.
We distinguish those three based on the URL structure (i.e., \texttt{i/communities/} or \texttt{/status/}). 
We find 1,271,698 users, 434,383 communities, and 1,447,668 unique post links. 
We similarly get user accounts and their post links in Truth Social, accounting for 372 users and 4,662 unique posts.

Telegram links are either public (i.e., groups where messages are public), private (i.e., groups that require invitations), or bots (i.e., programmable chatbots, typically trading bots). 
We differentiate private channels that start with a plus sign (``+'') after the domain and bots whose usernames ends with ``bot'' (case-insensitive)~\cite{tsuchiya2026large}. 
There are 800,441 public channels, 235,850 private channels, and 5,220 bots. 
We will use Telegram groups (both public and private) and Twitter communities to study community-based manipulation, and Twitter and Truth Social posts for potential post-based manipulation.

\noindent\textbf{Results: Community-based manipulation.}
We first focus on community-based manipulation and analyze their strategies. 
We show that most groups are associated with only a few coins and are short-lived, while some groups are highly active.
Table~\ref{tab:community_descriptive} illustrates the distribution of the number of coins each community has and the number of active days (i.e., days when they have at least one coin).
Most groups have 1~coin, and 95\% of groups only generate up to 5 coins (2 days). 
On the other hand, the top 0.1\% (1,478 groups, ranked by the number of coins created) have 60 or more coins constantly over time. 

\begin{table}[]
\caption{Community statistics: nr. of coins and active days.}
\begin{adjustbox}{width=\linewidth,center}
\begin{tabular}{@{}lrrrrrr@{}}
\toprule
Variable              & mean & 50\% & 95\% & 99\% & 99.9\% & max  \\ \midrule
Nr. of coins       & 1.91 & 1      & 5    & 13   & 60     & 7,630 \\
Nr. of active days & 1.29 & 1      & 2    & 6    & 21     & 346  \\ \bottomrule
\end{tabular}
\end{adjustbox}
\label{tab:community_descriptive}
\end{table}

\begin{figure*}
    \centering
    \begin{minipage}{0.19\linewidth}
        \centering
        \includegraphics[width=1\linewidth]{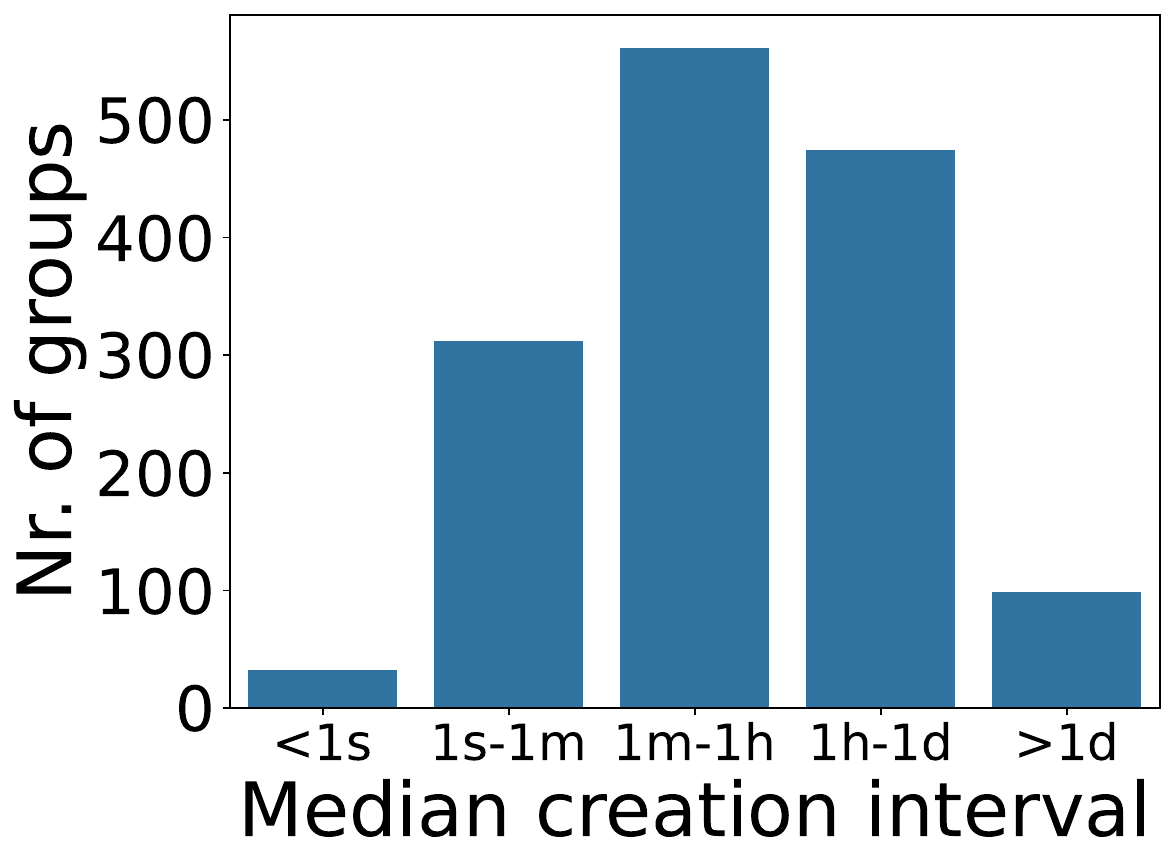}
        \phantomsubcaption
        \label{subfig:community_median_interval}
    \end{minipage}
    \hfill
    \begin{minipage}{0.19\linewidth}
        \centering
        \includegraphics[width=1\linewidth]{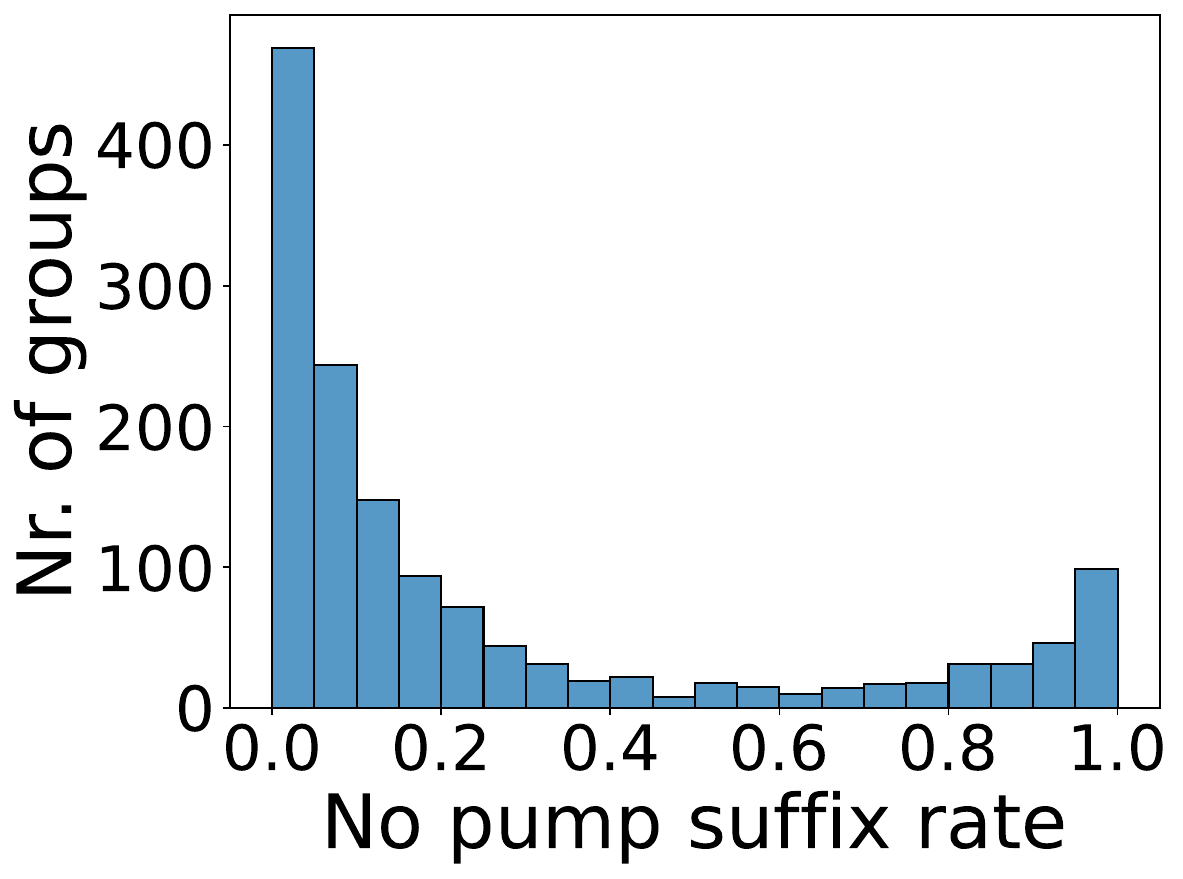}
        \phantomsubcaption
        \label{subfig:community_no_suffix}
    \end{minipage}
    \hfill
        \begin{minipage}{0.19\linewidth}
        \centering
        \includegraphics[width=1\linewidth]{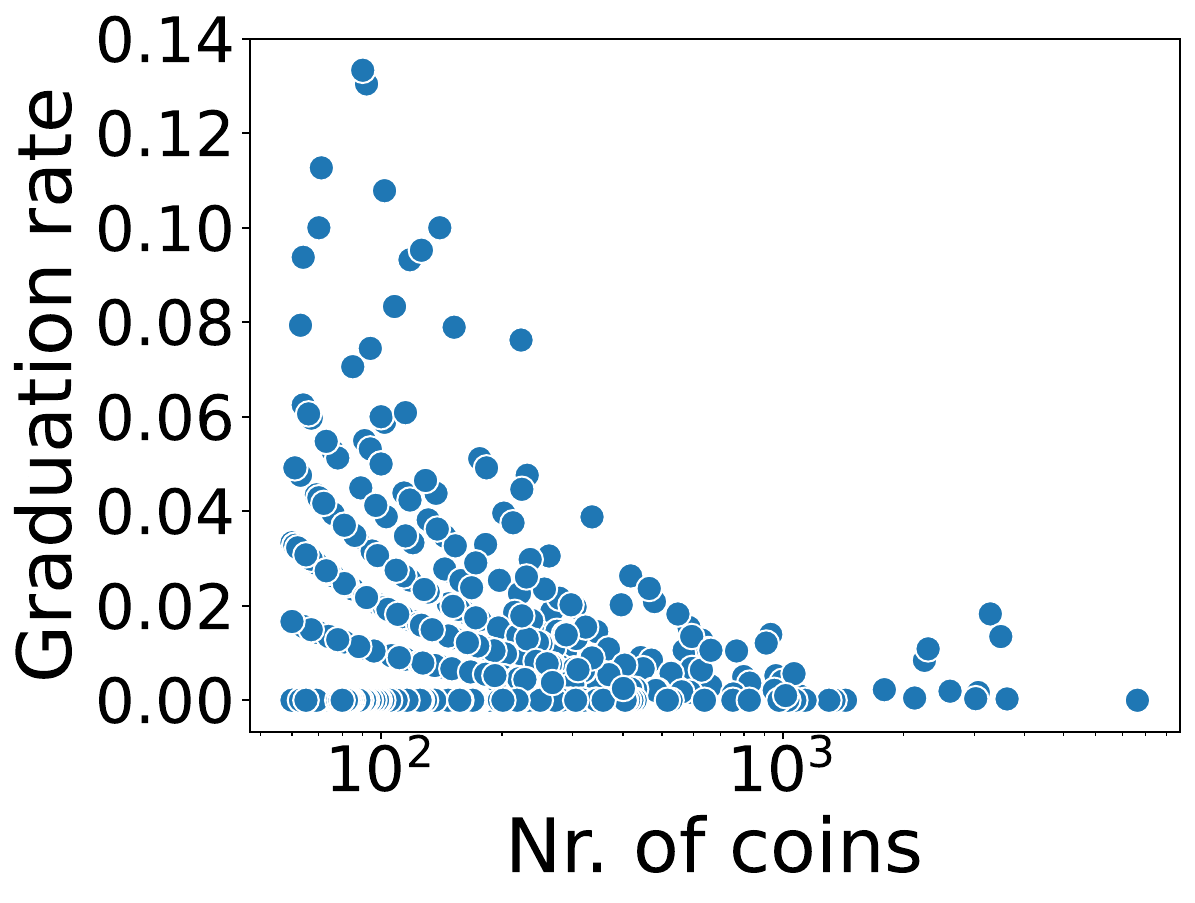}
        \phantomsubcaption
        \label{subfig:community_grad_rate_num_mints}
    \end{minipage}
    \hfill
    \begin{minipage}{0.19\linewidth}
        \centering
        \includegraphics[width=1\linewidth]{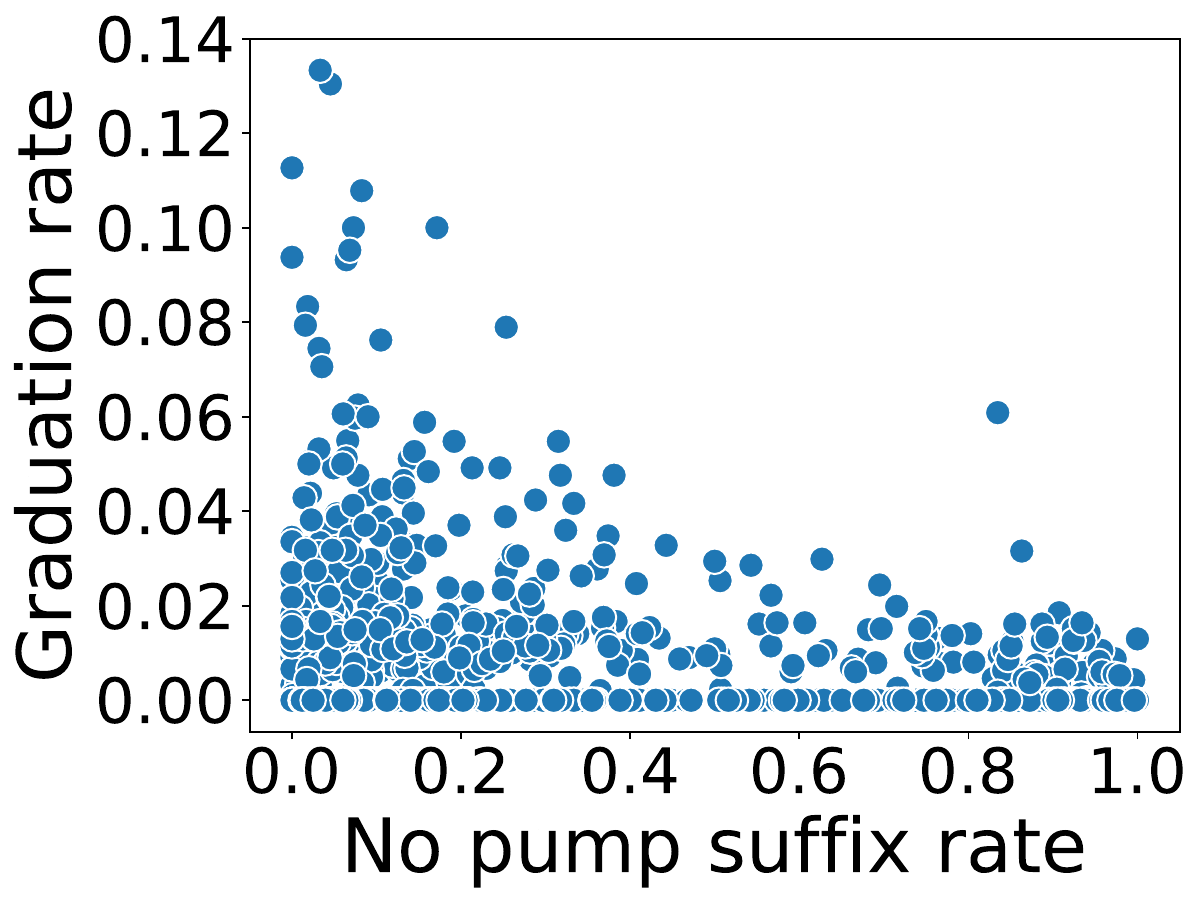}
        \phantomsubcaption
        \label{subfig:community_grad_rate_no_suffix}
    \end{minipage}
    \hfill
    \begin{minipage}{0.19\linewidth}
        \centering
        \includegraphics[width=1\linewidth]{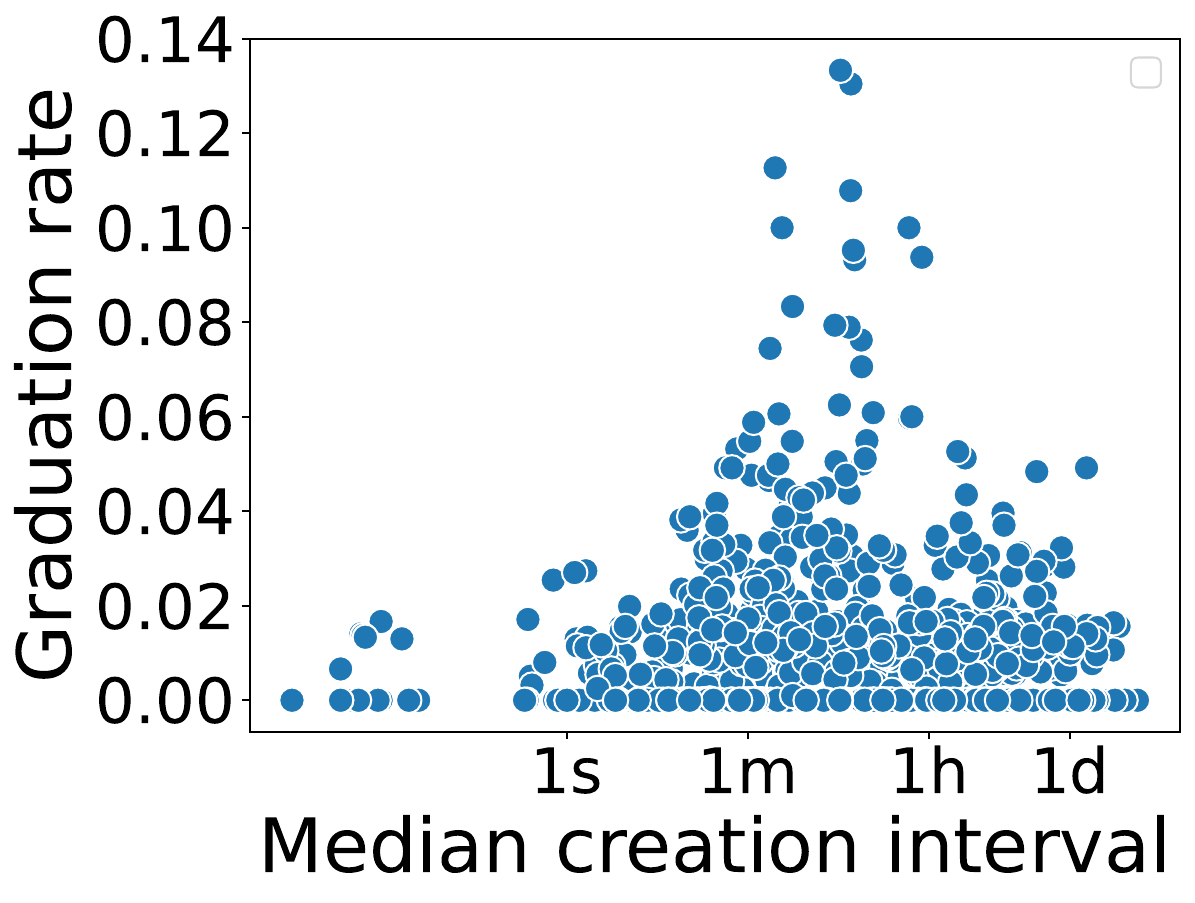}
        \phantomsubcaption
        \label{subfig:community_grad_rate_median_interval}
    \end{minipage}
    \caption{For each community, the distribution for (a) medium creation interval and (b) automation (``no suffix'') ratio. Each community's graduation rate vs (c) the number of coins, (d) automation ratio, and (e) medium coin creation interval.}
    \label{fig:community_stats}
\end{figure*}

We then turn our focus to these top 0.1\% of highly active groups. 
We observe variations in coin creation timing and automation level. 
Figure~\ref{subfig:community_median_interval} illustrates the distribution of the median coin creation interval between two consecutive coins generated by the same group; some groups generate a coin within a second, while others spend more than a day. 
Figure~\ref{subfig:community_no_suffix} shows the ``automation rate'' distribution, that is, the distribution of the ratio of coins without the \texttt{pump} suffix (likely to have been directly instantiated on the blockchain, through some automated mechanism, see \S\ref{sec:copycat}) over all coins created by the same group. 
In other words, when all coins have the \texttt{pump} suffix, $x=0$; when none of them do, $x=1$. 
The distribution is bimodal. 
Many channels appear to create coins via the website UI (low $x$), while others systematically interact directly with the blockchain ($x=1$).

We then examine these variables against the ``graduation rate,'' the number of graduated coins divided by the total number of coins created by each group, reflecting the  ``skillfulness'' of each group. 
We provide three scatter plots against the graduation rate: 
Fig.~\ref{subfig:community_grad_rate_num_mints} is the number of coins (i.e., the number of attempts), Fig.~\ref{subfig:community_grad_rate_no_suffix} is the automation (``no suffix'') rate, and \ref{subfig:community_grad_rate_median_interval} is the median creation interval. 
The graduation rate generally rises with fewer attempts, less automation, and a medium to longer creation interval. 
For instance, some groups with 100 attempts, no automation, and 30 min interval achieve nearly a 14\% graduation rate, 14 times higher than the baseline.
The groups appear to be more successful when they do not have too many coins, use the website UI, and do not spam coins in a short period of time. 
In Appendix~\ref{appendix:community_coin_creation_order}, we also show that the graduation rate drops significantly after the first coin, suggesting that continuing to create many coins in the same channel may not yield additional profits.

Finally, we qualitatively examine posts and messages from Telegram public groups and Twitter communities, document their activity level, and infer their coordination intent.
We look at the 20 most active communities (by number of coins) and the 20 most successful communities (by graduation rate). 
For each community, we manually review at least 50 messages and record the community activities (e.g., number of members, number of messages, presence of Solana addresses, or bots broadcasting coin information). 
The two samples reveal distinct patterns.
Communities with many coins tend to focus on trading and coin announcements rather than their underlying projects (e.g, channel owners and members hype upcoming coin releases). 
One notable scheme appears in Twitter communities (5 out of the 10 most active ones) that automatically generates a coin when someone joins; their average automation rate (i.e., no-pump suffix) is 96.1\%. 
In contrast, successful communities tend to post project announcements (with an occasional use of a bot to broadcast coin status), and have coins that are aligned with the project theme.  
An exception is an AI agent channel that randomly deploys coins after famous figures or characters. 
We documented the detailed method and results in Appendix~\ref{appendix:community_case_study}. 

\noindent\textbf{Results: Post-based manipulation.}
We next study post-based manipulation.
3,571,170 coins (23.5\% of all coins) are created (shortly) after 1,452,330 Twitter or Truth Social posts, highlighting the strong influence of these platforms on \pumpfun. 
We are interested in estimating a potential profit from creating a coin based on a specific post.  
We define the ``extractable value ($EV$)'' as the largest price difference multiplied by the amount of tokens the manipulator initially holds. 
The largest price difference is the highest daily price after the coin creation\footnote{On very rare occasions, \pumpfun appears to miscalculate the price, causing unrealistic (1,000,000$\times$) high price jumps. We exclude these data points by taking the maximum daily \emph{low} across the whole interval (i.e., the historical ``high'' not accounting for intra-day variations) as a conservative estimate.} minus the price of the coin creation in USD.
While manipulators can own up to 1~billion tokens (i.e., the maximum token supply for each coin), the more they own, the lower the returns they can expect since this depletes the coin supply available to potential buyers.
In this analysis, we set $x=0.1$ (i.e., the manipulator possesses 10\% of the initial share), though the number is easily adjustable.

We estimate $EV$ for all graduated coins\footnote{We assume $EV=0$ for non-graduated coins.} for each post and aggregate $EV$s if the post has more than one graduate coin. 
We focus on 12,544 posts (0.86\% of all posts) with a positive $EV$. 
Table~\ref{tab:post_descriptive} summarizes the distribution of the number of coins (including non-graduates) and the total of $EV$ per post. 
The median post has 12 coins, and a total $EV$ of 135~USD.
We do not find any significant correlation between the number of coins and $EV$. 
31 posts (the top 0.002\% of all posts) have $EV\geq 1$ million USD, suggesting that the post-based manipulation can be highly competitive but lucrative.

\begin{table}[]
\caption{Post statistics: nr. of coins and the total $EV$ in USD.}
\begin{adjustbox}{width=\linewidth,center}
\begin{tabular}{@{}lrrrrrr@{}}
\toprule
Variable      & mean & 50\%   & 95\%    & 99\%      & 99.9\%     & max            \\ \midrule
Nr. of coins  & 22.52 & 12     & 76      & 149       & 338        & 3,406          \\
Total $EV$ (USD) & 28,260 & 135 & 12,748 & 126,445 & 3,683,136 & 113,633,854 \\ \bottomrule
\end{tabular}
\end{adjustbox}
\label{tab:post_descriptive}
\end{table}

To understand how social media posts generate profits, we manually review the 31~posts with $EV\geq 1$M~USD.
These posts together account for 83.4\% of the total $EV$ across all posts. 
For each post, we record engagement statistics (e.g., number of comments, reposts, likes, and number of followers the creator has), and whether the post (and the original post if the post is a reply or repost) includes pictures or videos. 
The complete results are in Appendix~\ref{appendix:social_media_posts}. 

These profitable posts tend to come from high-profile accounts.
The post creators have a median of 42,600 followers, ranging from 5,000 to 240~million followers. 
Their posts are highly engaging, with an average of 1,001 comments, 4,053 reposts, and 20,246 likes, and a median of 56 comments, 93 reposts, and 413 likes.
These numbers are an order of magnitude higher than those of average Twitter posts: 2.6 comments, 6.7 reposts, 32.9 likes~\cite{statista_twitter}. 

We also find that these successful posts generally fall into one of three categories. 
Inspired by Long et al.~\cite{long2025bridging}, we classify these posts into Culture, News, and Animal. 
Culture posts ($n=13$) typically feature internet jokes often involving AI or celebrities (e.g., Elon Musk). 
News posts ($n=8$) generally refer to announcements or real-world events (e.g., Ross Ulbricht's pardon). 
Animal posts ($n=5$) often introduce pets or AI-themed characters. 
20 out of 27 posts contain pictures or videos, suggesting that visual content is an important component.

As a case study, we choose five notable posts (out of 31): \path{truth_terminal}, \texttt{bitcoinmagazine}, \texttt{d33v33d0}, \texttt{tong0x}, \texttt{tonyplasencia3} to illustrate 1) the content of each post and 2) how it led to the creation of its most profitable coin. 
In all five cases, the original posters did not seem to have created the coin;
posters may have no malicious intent, yet others can still profit from the post. 
Some posters endorse the token (e.g., \path{truth_terminal}) or receive some coin shares (e.g., \texttt{d33v33d0}).
In contrast, other posters explicitly disassociate from the coin by 1) replying within the post's thread to state that the coin does not belong to the poster (e.g., \texttt{tong0x}, \texttt{tonyplasencia3}), 2) stating in the post itself that the poster has no intention of creating a meme coin. 
The time from post to coin creation varies widely: from 4 minutes (a news announcement, \texttt{bitcoinmagazine}) to a few months (a dog picture, \texttt{tonyplasencia3}). 
Details for each coin case study, along with the snapshots of the post, are in Appendix~\ref{appendix:post_case_study}.

 \section{Market-manipulation-as-a-service}
\label{sec:mmaas}
In this section, we identify and investigate a set of websites and software tools that \textit{could} enable users to perform the above manipulations with ease: Market-Manipulation-as-a-Service (MMaaS).

\subsection{MMaaS websites}
\label{subsec:mmaas_websites}
We first identify websites or apps with complex feature support and sophistication, e.g., those that require access to blockchain nodes or social media feeds with low latency. 
As described in \S\ref{tab:copycat_detection}, \pumpfun typically hosts coin images on IPFS.
However, some external tools for launching coins host images on their own infrastructure, which we can use to identify a lower bound on potential MMaaS activity related to \pumpfun.

We follow the same process as \S\ref{subsec:external_platforms} to extract (non-IPFS) domains from the coin metadata and manually infer potential websites from these domains (e.g., [platform\_x] from [metadata.platform\_x.com]). 
We look at the top 30 domains by frequency and only include those that 1) are currently available and identifiable (e.g., some are in cloud services, such as Deutsche Telekom, Digital Ocean), 2) do not require authentication or registration,
and 3) provide at least two functionalities defined below. 
These criteria reduce the list to four distinct services. 

For each website, we check whether it \textit{claims} to 1) provide low-latency services, 2) create multiple addresses (e.g., to buy coins in one transaction ``bundling''), 3) copy existing coins (i.e., often called ``vamping''), and 4) track social media feeds in real time. 
Importantly, we did \emph{not} purchase or use these services to validate their claimed functionality for ethical reasons in Appendix~\ref{sec:ethics}.
Two authors annotated all four services on Apr. 14--15, 2026, following the annotation codebook (in Appendix~\ref{appendix:mmaas_codebook}).
Both authors agreed on 15 out of 16 cases (4 functionalities across 4 services).  
They met up to resolve the one disagreement. 
We summarize our results below and leave individual annotation results in Appendix~\ref{appendix:mmaas_codebook}.

All services provide low-latency features, while some services help create multiple addresses, copycat coins, and track social media in real time, which corroborates our findings in \S\ref{sec:evaluation_clustering}, \S\ref{sec:copycat}, and \S\ref{subsec:external_platforms}, respectively. 
Furthermore, one service allows users to create multiple addresses from a mixer (to obfuscate an identity potentially), and another service enables users to generate a vanity contract address ending with ``pump'' (to imitate coin creation on the \pumpfun UI potentially).

\subsection{MMaaS software}
\label{subsec:mmaas_software}
We next investigate open-source software that primarily supports simple automation (e.g., software that does not require access to blockchain nodes, for instance, software used to post comments automatically). 
We run the keyword search  ``pump fun comment bot'' on GitHub, which identifies 29 repositories. 
We manually investigate all projects and exclude obvious duplicates and those without enough information in the \texttt{README} file. 
In total, we confirm 14 repositories \emph{advertise} \pumpfun comment bots. 
Similar to He et al.~\cite{he2026six}, we look at the \texttt{README} file, since the repositories often 1) redact some files or 2) redirect to external links. 
To avoid engaging in malicious activity ourselves, we cannot run the software to verify its advertised functionality.

We follow the same procedure as above -- for the website analysis -- to annotate software functionalities. 
We check whether the repository \textit{advertises} to 1) create fresh wallets for commenting, 2) use both fresh wallets and coin creator wallets for commenting (for a more genuine look), 3) provide predefined comment sets, 4) use AI to generate user profiles or comments, and 5) deploy anti-detection (e.g., proxies, CAPTCHA solvers). 
The two authors agree on 57 out of 70 questions (5 functionalities, 14 services). 
The major discrepancy comes from 1) ambiguity in the \texttt{README} file (e.g., whether the services or users prepare the predefined comments), 2) vague codebook definitions (e.g., terminology confusion), and 3) simple mistakes by one of the authors.  
If the description is unclear, we do not select that functionality (conservatively). 
Our final annotation results and the codebook can be found in Appendix~\ref{appendix:mmaas_codebook}.

Overall, at least 8 repositories create fresh wallets (i.e., accounts) for users, while 4 repositories mix fresh accounts with developer or creator accounts to create even more natural social interaction (often called ``shill'' mode). 
6 repositories provide a predefined set of comments for easier execution; the \texttt{README} file often claims that the software owner handpicked comments from the wild. 
On the other hand, two repositories use AI to generate comments or fake profiles (e.g., using OpenAI). 
10~repositories have anti-detection tools such as CAPTCHA solving or proxy rotations (e.g., when creating accounts and posting comments on \pumpfun).

With these services, users can execute manipulation without any technical knowledge or infrastructure. 
They can easily create multiple blockchain addresses to obfuscate token ownership, copy existing tokens, and monitor social media feeds in real time with low latency.  
Users can also create realistic social interactions (e.g., a coin developer talking to other accounts) through custom or AI-generated comments.
These tools can even circumvent platform defenses (e.g., CAPTCHAs) or network-based defenses by hiding their IP addresses behind proxies.
The websites and GitHub repositories we identify are publicly accessible: we may find more of these services in Telegram channels or underground marketplaces, which are considerably harder to find. 
 \section{Discussion and Conclusion}
\label{sec:discussion}

Our findings have two clear takeaways.
First, \textbf{market manipulation activities are widespread, even under conservative estimates}. 
Our analysis shows that at least 17\% of all trading transactions are wash trades, and at least 10\% of all coins on \pumpfun are copycats. 
Several manipulation strategies are demonstrably effective: 
coins with high levels of wash trading activity are more likely to succeed. 
Similarly, some social media groups tend to have higher success rates. 
While other strategies, such as copycat creation, appear less effective on average, they remain important components of broader manipulation campaigns designed to mislead others.

Second, \textbf{manipulation strategies are sophisticated}. 
We have consistently observed that strategic actors deploy custom smart contracts that directly interact with the Solana blockchain in a highly automated manner. 
For instance, manipulators frequently bundle multiple trades within a single transaction (WT1 in \S\ref{wash_trading_section} or DP1 in \S\ref{sec:evaluation_dumping}), which is not possible through the \pumpfun web interface. 
Similarly, we find that some copycat creators (in \S\ref{sec:copycat}) and Telegram or Twitter groups (in \S\ref{subsec:external_platforms}) consistently generate coins outside the \pumpfun web interface (i.e., no pump suffix).
The existence of Manipulation-as-a-Service (MMaaS) tools (\S\ref{sec:mmaas}) further facilitates the execution of sophisticated manipulation strategies.

As other competitors follow similar web interfaces and on-chain implementations, we believe most of the manipulation strategies here are also applicable to them.
We leave the investigation of such manipulations' prevalence on other platforms to future work.

\noindent\textbf{Mitigations. }
Our findings suggest that strategic actors on \pumpfun manipulate a variety of signals, including trading volume, price charts, and comments.
These strategies highlight the need for interventions that both raise costs for manipulators and help users better interpret the signals they observe. 
We outline interventions along two dimensions: technical and regulatory interventions.

\emph{Technical interventions.}
Overall, we believe there is a need for better quality and integrity of displayed signals. 
Platforms in the launchpad ecosystem, such as \pumpfun, wallets, chain scanners, and statistics aggregators can detect or even suppress manipulative activity such as wash trading. 
Such detection can increase the friction and cost for manipulators and make it harder to mislead users. 
For example, atomic wash trading is easy to detect and unlikely to contain false positives. 
While adversaries may adapt to evade detection, doing so increases their transaction costs. 

In a similar vein, platforms should expose more robust provenance signals. 
For instance, they can create the funding graph (in \S\ref{sec:evaluation_clustering}) for each coin creator and show if a creator uses multiple addresses for obfuscation.   
Moreover, \pumpfun can also flag coins whose addresses do not end with the ``pump'' suffix (i.e., likely creating coins outside \pumpfun's UI), which suggests deviation from the typical coin creation process and may deserve more scrutiny.
Additionally, \pumpfun can show the number of copycat coins, e.g., ``this coin has $x$ coins with the same name, symbol, description, and image.''
This additional information would help prevent users from accidentally buying the wrong coin. 
Simultaneously, we advise users to adjust their expectations about simple indicators such as trading count or coin comments that are easy to manipulate.

\emph{Regulatory interventions.}
Regulatory bodies, such as the U.S. Securities and Exchange Commission (SEC) or its counterparts in other jurisdictions, can also play a role in helping address manipulation in the ecosystem. 
For example, they can create a regulatory framework to account for new forms of manipulation. 
We show that social media influencers \textit{could} tweet and create a coin simultaneously to extract millions of profits on \pumpfun potentially---this may be very similar in spirit to insider trading. 
Regulators can also consider taking down MMaaS websites or coordinate with GitHub moderators to remove repositories that enable manipulation at scale. 
However, software-level interventions may be less effective given the ease of creating new websites and repositories.

 \section*{Acknowledgement}
\label{sec:ack}
We thank Bryan Routledge, Ariel Zetlin-Jones, Viraj Shah and Kamya Singh for research discussions, review of relevant literature and assistance with preliminary data collection. 
This work was supported by the Cylab Presidential Fellowship, CyLab  Security  and  Privacy  Institute  seed  funding grant, the Hasler Foundation, Nakajima Foundation, and Google Academic Research Award.
 
\bibliographystyle{ACM-Reference-Format}
\bibliography{reference}

\appendix \section{Open Science} \label{sec:open_science}

We release the blockchain data and the scripts needed to reproduce our analysis
of trade manipulation. Our data includes coin minting transactions (e.g.,
the coin's first transaction) for all 15 million coins, as well as the
entire transaction history for both samples: the \onepctsample and the \fivedaysample.

This consists of a large (approximately 2~TB
uncompressed) Postgres database, available at \artifactURL. 
The page also contains a set of scripts to
help interact with the database.

We cannot release the data we collected from the \pumpfun website
because \pumpfun's Terms of Use (ToU) prohibits data (re)distribution.

\section{Ethical Considerations}
\label{sec:ethics}

This section discusses the ethics of our data collection and the potential impact of publishing our results. 

\noindent \textbf{Blockchain data.}
We collected the blockchain transaction data from the Solana blockchain using third-party RPC service providers such as Chainstack and Syndica.
This blockchain data is publicly accessible to everyone.
We also use Arkham Intelligence's API~\cite{arkham_api_guide} data to supplement wallet information. 
As the Solana blockchain wallets are pseudonymous (i.e., base58 strings), we do not consider them as Personally Identifiable Information (PII). 

\noindent \textbf{\pumpfun website data.}
We collected coin information, the coin comments, and the price data from the \pumpfun website.
These data are critical to identify new classes of attacks in \S\ref{sec:copycat}, \S\ref{sec:social_manipulation}, and \S\ref{sec:mmaas}. 
We only use \pumpfun's APIs~\cite{bankk2026pumpfun_api} (no web scraping), following their ToS at the time of data collection. 
\pumpfun did not provide \texttt{robots.txt} at that time.
To minimize load, we leave at least 4--5 seconds between each request.
For instance, when we retrieve coin information for 15.2 million coins, we have made about 304,000 requests (i.e., 50 coins for each query) and spread them across 25 days. 
We did not scrape web pages or download media files (e.g., images and videos).  
\pumpfun accounts are also blockchain wallets (or occasionally user-defined usernames), which are not considered personally identifiable information (PII).
Therefore, the IRB at our institution(s) do not consider this human subject research.

\noindent \textbf{MMaaS.}
We investigated the potential websites or software tools that could facilitate the execution of market manipulation.  
We only view the website or the GitHub page, but do not interact with these services or execute their software to avoid supporting their services or risk installing malicious code. 
We opted not to report these services for the reasons below. 
First, some features have benign uses. 
For instance, users can create multiple blockchain addresses for privacy-preserving purposes. 
Second, we could not verify what these services actually do without using them or executing their software. 
Accordingly, we focus on documenting what they \textit{advertise} and discuss implications for regulators in \S\ref{sec:discussion}. 
When reporting our qualitative analysis in this paper, we anonymize their service names (e.g., Service $A$) to prevent misuse. 

\noindent \textbf{Social impact} 
Malicious actors could misuse our findings and refine their attack strategies to prey on inexperienced traders.  
Simultaneously, our work helps inexperienced traders to recognize manipulation schemes and take precautionary measures. 
As discussed in \S\ref{sec:discussion}, our findings could also inspire defenses for many actors in this space, including chain scanners, wallet providers, and financial regulators. 
Following the utilitarian ethics advocated in cryptomarket research~\cite{martin2016ethics}, we believe 
the advantages of measuring the \pumpfun ecosystem and reporting our findings outweigh the potential harms.
As a part of responsible disclosure, we reported and discussed our findings with \pumpfun team.

\section{Generative AI Usage}
We confirm that the authors wrote all the text in this paper. 
We used AI-based tools (e.g., Grammarly, Claude, Copilot) solely for grammar and spell-checks and to improve the clarity of the author-written text. (As is common practice, most text was rewritten multiple times by multiple authors anyway, bearing very little resemblance to the first draft.)
We sometimes used Copilot for automatic code completion (primarily for figures) to reduce typing errors. 
However, we manually verified and double-checked every single line of code.
 \section{\pumpfun trading fees} 
\label{appendix_pump_fun_trading_fees}
This section explains the transaction fees on \pumpfun.
When users buy or sell tokens on \pumpfun, they pay transaction fees to both \pumpfun and a coin creator based on transaction volume (after May 2025)~\cite{pumpfun2025fees_graduation}.
The fee amount also depends on the coin status (pre- and post-graduation) and the market cap.
The maximum fee has been 1.25\% since the launch of \pumpfun.
Wash trading detection heuristic \textbf{WT2} matches the buy and the sell trade if their trading volumes are equivalent. 
However, due to transaction fees, the amount of SOL users receive back is smaller than their initial capital.  
To account for this, we allow a tolerance of transaction volume---1.25\% per trade, and the blockchain fees of 0.000015 SOL (i.e., 15,000 lamports).
 \section{Wash trading detection for \fivedaysample} 
\label{appendix-5-day-sample-wash-trading}
In \S\ref{wash_trading_section}, we primarily report our wash trading detection results on \onepctsample. 
This supplementary section describes our detection results for \fivedaysample.
Under WT1, we find 1,804,378 wash trading transactions, which happen in \PctSelectedHOneWashTokens~of coins (i.e., coins with at least one wash trading transaction). 
Under WT2, at the time window size of 1 second, only 10.94\% of coins have wash trading transactions.
If we increase the window size, the ratio of coins flagged by WT2 dramatically increases, reaching 88.01\% percent with a 1-day window, again, hinting at the presence of false positives.

\begin{figure*}[t]
    \centering

    \begin{subfigure}[t]{0.48\textwidth}
        \centering
        \includegraphics[width=\linewidth]{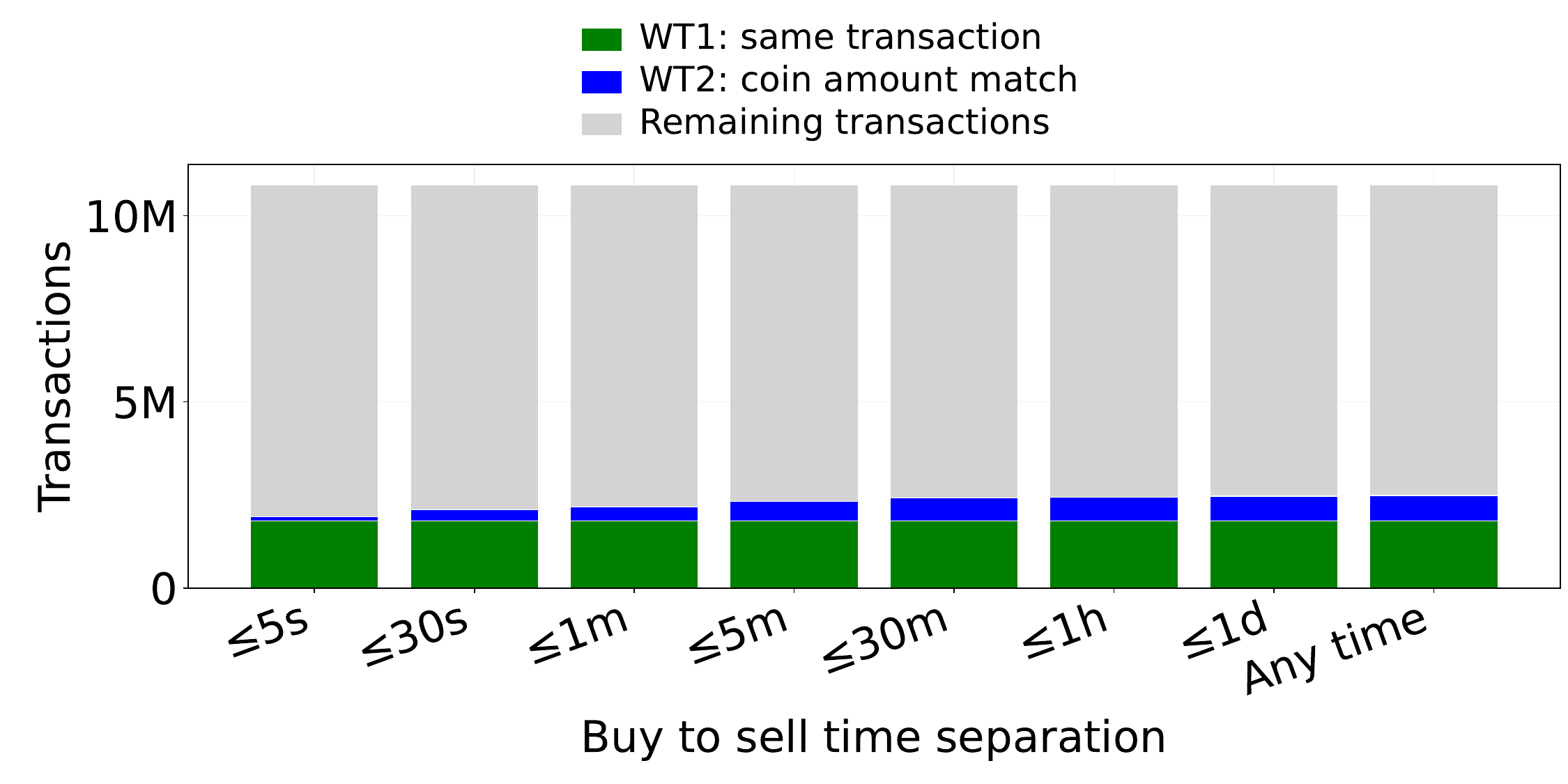}
    \end{subfigure}\hfill
    \begin{subfigure}[t]{0.48\textwidth}
        \centering
        \includegraphics[width=\linewidth]{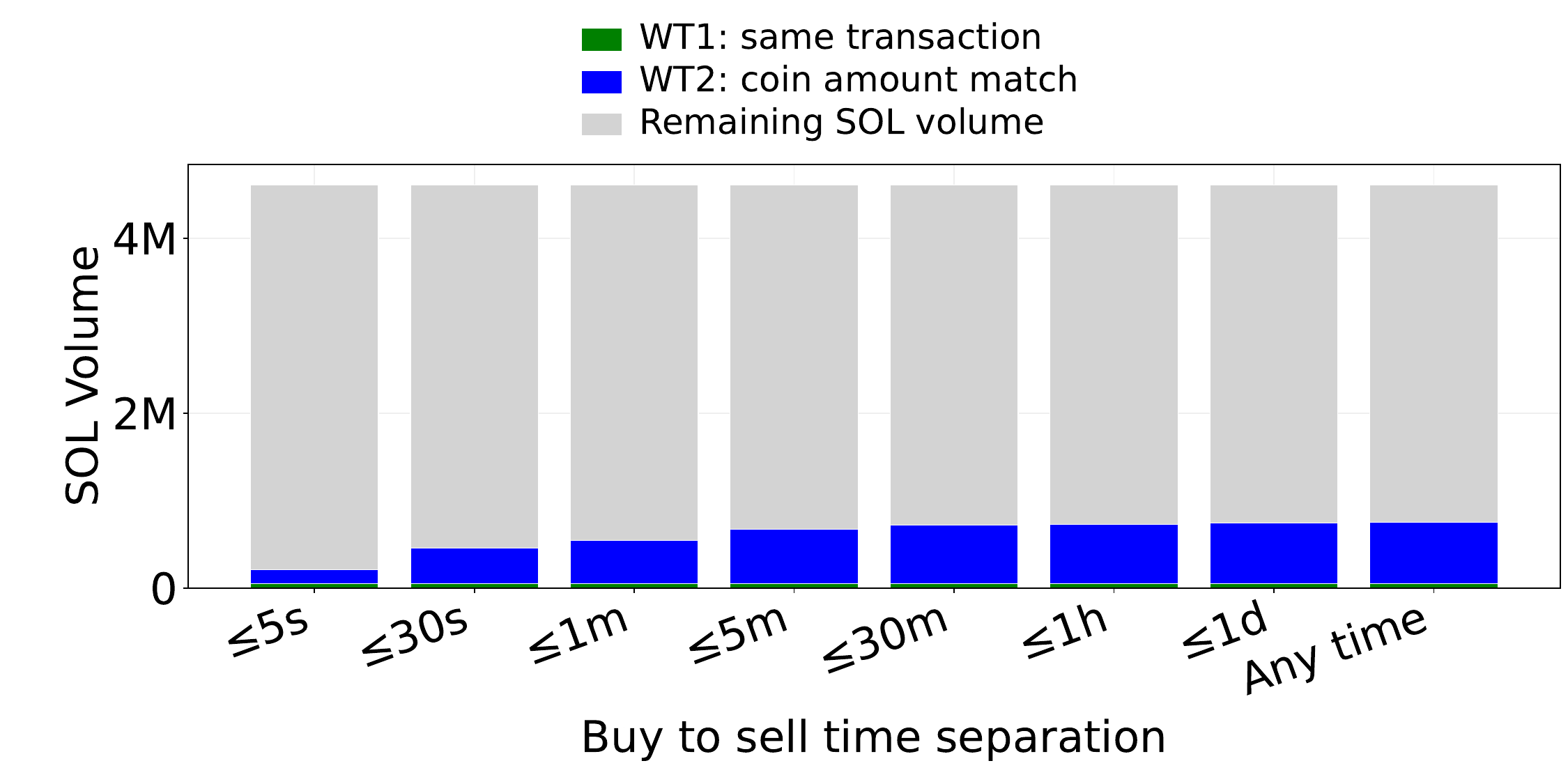}
    \end{subfigure}

     \caption{Wash trading transaction count (left) and volume in SOL (right) under WT1 and WT2 for \fivedaysample}
    \label{appendix_wash_trading_heuristics}
\end{figure*}

The results for \fivedaysample are consistent with those for \onepctsample. 
We list the figures and tables for reference.
Figure~\ref{appendix_wash_trading_heuristics} illustrates the number of wash trading counts and the volume. 
Table~\ref{share_of_coins_flagged_by_wt1_5_day_sample} illustrates the relationship between the ratio of wash trading transactions and the total number of transactions for each coin.
Table~\ref{graduation_by_number_of_transactions_matching_wt1_5_day_sample} categorizes coins based on the number of wash trading transactions and shows their graduation rate. 

\begin{table}
\centering
    \caption{Nr. of total transactions and the wash trading ratio for the \fivedaysample.}
    \begin{tabular*}{\columnwidth}{@{\extracolsep{\fill}}lrr@{}}
    \toprule
    No. of txs & Wash trading ratio (\%) & Nr. of coins \\
    \midrule
    1--10         & \PctSelectedMeanWashShareTxOneToTen & \NSelectedMeanWashShareTxOneToTen \\
    11--100       & \PctSelectedMeanWashShareTxElevenToHundred & \NSelectedMeanWashShareTxElevenToHundred \\
    101--1,000    & \PctSelectedMeanWashShareTxHundredOneToThousand & \NSelectedMeanWashShareTxHundredOneToThousand \\
    1,001--10,000 & \PctSelectedMeanWashShareTxThousandOneToTenThousand & \NSelectedMeanWashShareTxThousandOneToTenThousand \\
    10,001+       & \PctSelectedMeanWashShareTxTenThousandOneToHundredThousand & \NSelectedMeanWashShareTxTenThousandOneToHundredThousand \\
    \bottomrule
    \end{tabular*}
    \label{share_of_coins_flagged_by_wt1_5_day_sample}
\end{table}

\begin{table}
    \centering
    \caption{Graduation rates by number of wash trading transactions for the \fivedaysample.}
    \begin{tabular*}{\columnwidth}{@{\extracolsep{\fill}}lrr@{}}
        \toprule
        Nr. of wash trading txs & Grad rate (\%) & Nr. of coins \\
        \midrule
        0        & \PctSelectedGradCohortZero & \NSelectedGradCohortZero \\
        1--5     & \PctSelectedGradCohortOneToFive & \NSelectedGradCohortOneToFive \\
        6--25    & \PctSelectedGradCohortSixToTwentyFive & \NSelectedGradCohortSixToTwentyFive \\
        26--100  & \PctSelectedGradCohortTwentySixToHundred & \NSelectedGradCohortTwentySixToHundred \\
        101--500 & \PctSelectedGradCohortHundredOneToFiveHundred & \NSelectedGradCohortHundredOneToFiveHundred \\
        500+     & \PctSelectedGradCohortFiveHundredPlus & \NSelectedGradCohortFiveHundredPlus \\
        \bottomrule
    \end{tabular*}
    \label{graduation_by_number_of_transactions_matching_wt1_5_day_sample}
\end{table}

 \section{Creator address clustering} 
\label{appendix_creator_address_obfuscation}

\begin{table}[t]
\centering
\footnotesize
\setlength{\tabcolsep}{4pt}
\caption{Multi-address cluster summary.}
\label{tab:clustering_summary}
\resizebox{\columnwidth}{!}{\begin{tabular}{lrrr}
\toprule
Hops & Multi-address Clusters & Address \% & Created coin share \%  \\
\midrule
1-hop & 322,620  & 48.02\% & 62.99\% \\
2-hop & 295,978 & 53.23\% & 67.54\% \\
3-hop & 290,319 & 54.98\% & 68.78\%  \\
\bottomrule
\end{tabular}}
\end{table}

In \S~\ref{address_clustering_description}, we introduce our coin creator clustering approach.
We demonstrate that clusters created with the information deeper into the clustering graph connect more addresses together and better show the common coin creation.
This supplementary section contains Table~\ref{tab:clustering_summary}, which includes the count of multi-address clusters, their proportion of coin creators in multi-address clusters to all coin creators, and the ratio of coins created by multi-address clusters, along with the number of hops in the common funding graph.
 \section{Coordinated sell in \fivedaysample}
\label{appendix-5-day-sample-dumping}

\begin{figure}[t]
    \centering
    \includegraphics[width=\linewidth]{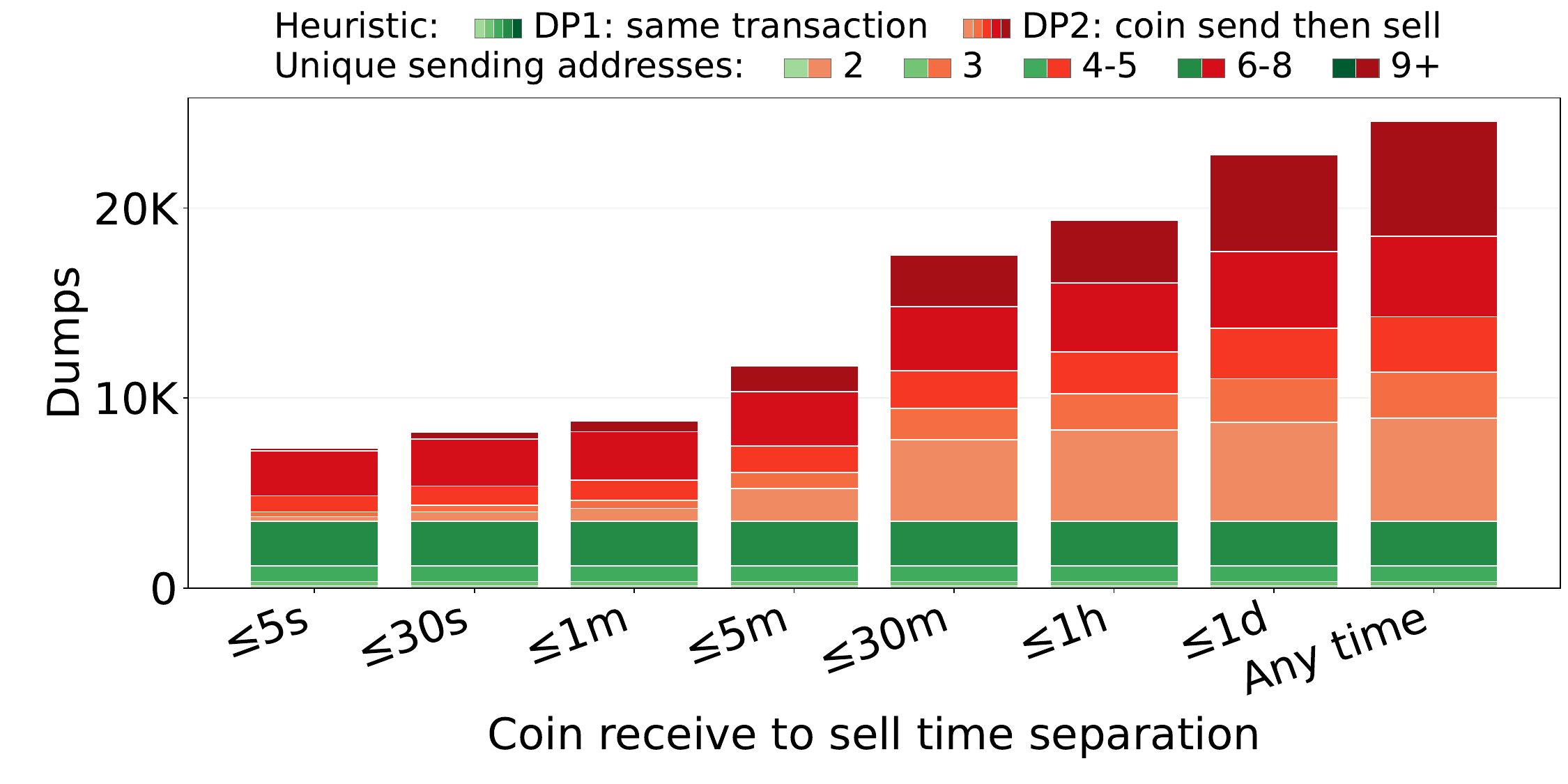}
    \caption{Nr. of dump instances for DP1 and DP2 in \fivedaysample.}
        \label{5day_sample_detected_dump_counts}
\end{figure}

\begin{figure}[t]
    \centering
    \includegraphics[width=\linewidth]{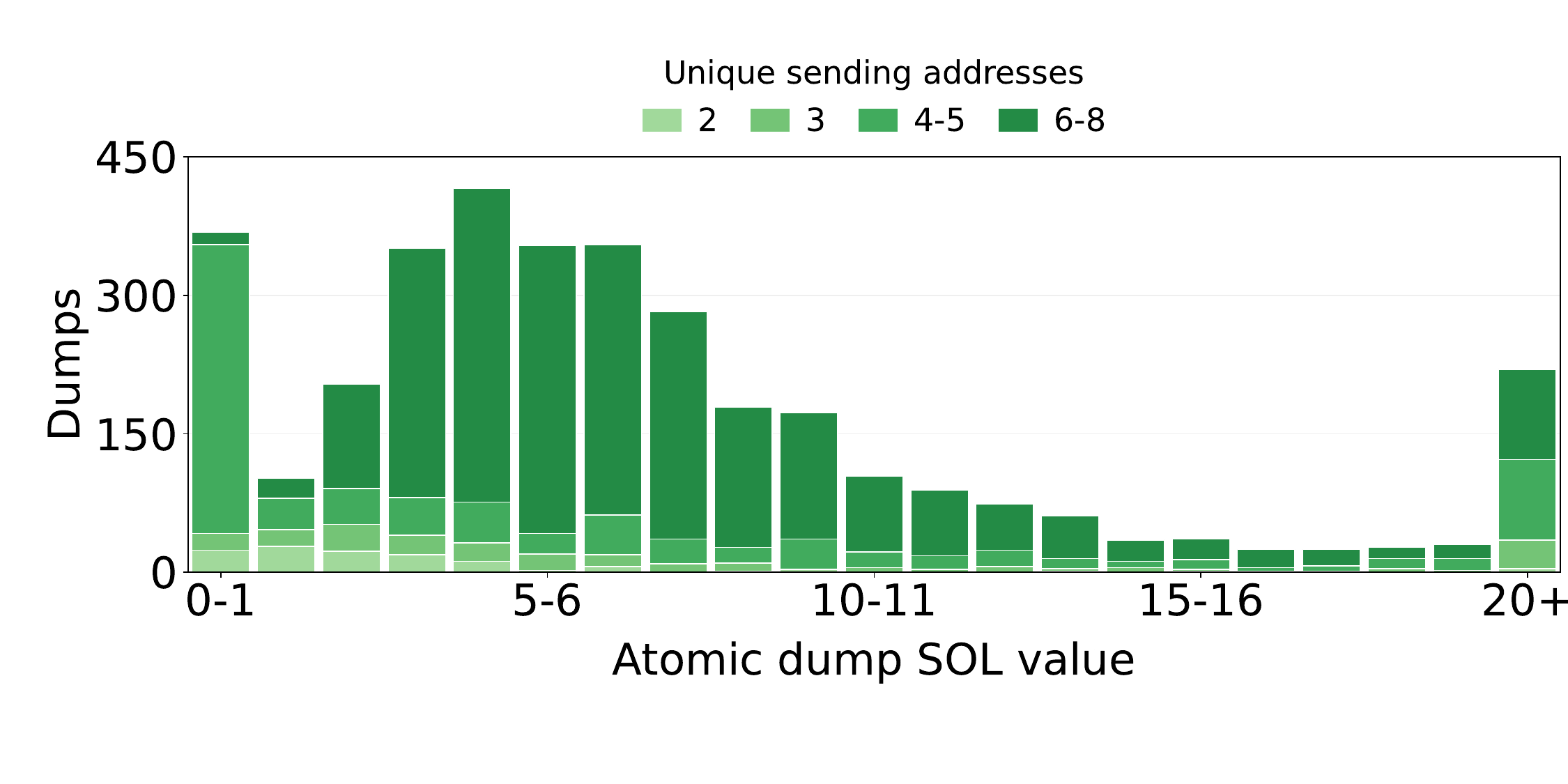}
    \vspace{-20pt}
    \caption{Distributions of dump transaction volume (in SOL) for \fivedaysample.
}
        \label{5day_sample_received_sol_distribution}
\end{figure}

In \S\ref{sec:evaluation_dumping}, we primarily report our coordinated sell detection results on \onepctsample. 
This supplementary section describes our detection results for \fivedaysample.
Under DP1, we find 3,510 cases of a coordinated sell. 
Under DP2, at the time window sizes of 5 seconds, 1 hour, and 1 day, we find 7,351, 55,354, and 74,635 coordinated sell events.
The other results for \fivedaysample are also consistent with those for \onepctsample. 
Figure~\ref{5day_sample_detected_dump_counts} illustrates the number of dump instances flagged by DP1 and DP2 with different time windows.
Figure~\ref{5day_sample_received_sol_distribution} shows the distributions of dump transaction volume (in SOL), flagged by DP1.

 \section{Clone coins} 
\label{appendix:clone_coins}

This supplementary section covers coins that share the same metadata (Heuristics 1: coin name, symbol, description, image) and are created by the same creator or belong to the same creator cluster.  
We call them ``clone'' coins. 
We identify 1,125,361 clone coins and the corresponding 440,162 original coins. 
We find that clone coins exhibit characteristics similar to copycats. 
Original coins have a higher graduation rate (1.03\%, 4,524 coins), compared to clones (0.66\%, 7,400 coins). 
They appear to be less automated, having a much lower no-pump suffix rate (19.20\%), compared to the clones (30.37\%).  
Yet, compared to originals of copycats (in \S\ref{sec:copycat}), originals of clones have a significantly lower graduation rate and are more automated.
This implies that owners typically deploy a clone when the original coin fails or automatically launch many coins at once in the hope that one gains traction. 

Just like copycats, each clone group (i.e., the original and the clone coins) rarely produces multiple graduated coins. 
The number of clone groups with more than 1 graduated coin is 1,049 (0.24\%). 
In terms of the coin creation timing, graduation probability is slightly higher for the earlier coins, but this effect is less clear, given the significantly smaller number of graduated coins for later coins. 
 \section{External platform count} 
\label{appendix:external_platform_count}
This appendix explains the external platforms linked to each coin (in \S\ref{subsec:external_platforms}). 
Table~\ref{tab:top10_platforms_open_fields} counts the number of coins (no deduplications) that have links to the top ten platforms for three open fields in the coin's profile page: ``\texttt{twitter},'' ``\texttt{telegram},'' and ``\texttt{website}.'' 
The external platforms linked give clues about the type of meme coin involved: community-based (e.g., Telegram groups, Twitter communities), content-based (e.g., Twitter, Instagram, or Reddit posts), video-based (e.g., Twitch, YouTube, or TikTok videos), or concept-based (e.g., Wikipedia, GitHub).

\begin{table}[]
  \caption{The top ten platforms by the number of coins linked by their three open fields: \texttt{twitter}, \texttt{telegram}, \texttt{website}}
\begin{adjustbox}{width=0.7\linewidth,center}
\begin{tabular}{@{}lrrr@{}}
\toprule
platform $\backslash$ field     & \texttt{twitter}   & \texttt{telegram}  & \texttt{website}   \\ \midrule
Twitter      & 8,320,474 & 432,441   & 1,505,489 \\
Telegram     & 164,012   & 2,157,025 & 167,484   \\
TikTok       & 40,956    & 31,799    & 288,586   \\
Instagram    & 40,235    & 29,663    & 186,595   \\
YouTube      & 38,496    & 30,513    & 284,971   \\
Reddit       & 18,721    & 15,289    & 97,213    \\
Wikipedia    & 16,918    & 13,952    & 79,971    \\
Twitch       & 8,818     & 7,027     & 41,989    \\
GitHub       & 8,104     & 9,898     & 63,681    \\
TruthSocial & 8,076     & 991       & 23,609    \\ \bottomrule
\end{tabular}
\end{adjustbox}
\label{tab:top10_platforms_open_fields}
\end{table}
 \section{Community coin creation order} 
\label{appendix:community_coin_creation_order}

\begin{figure}
    \centering
    \includegraphics[width=0.8\linewidth]{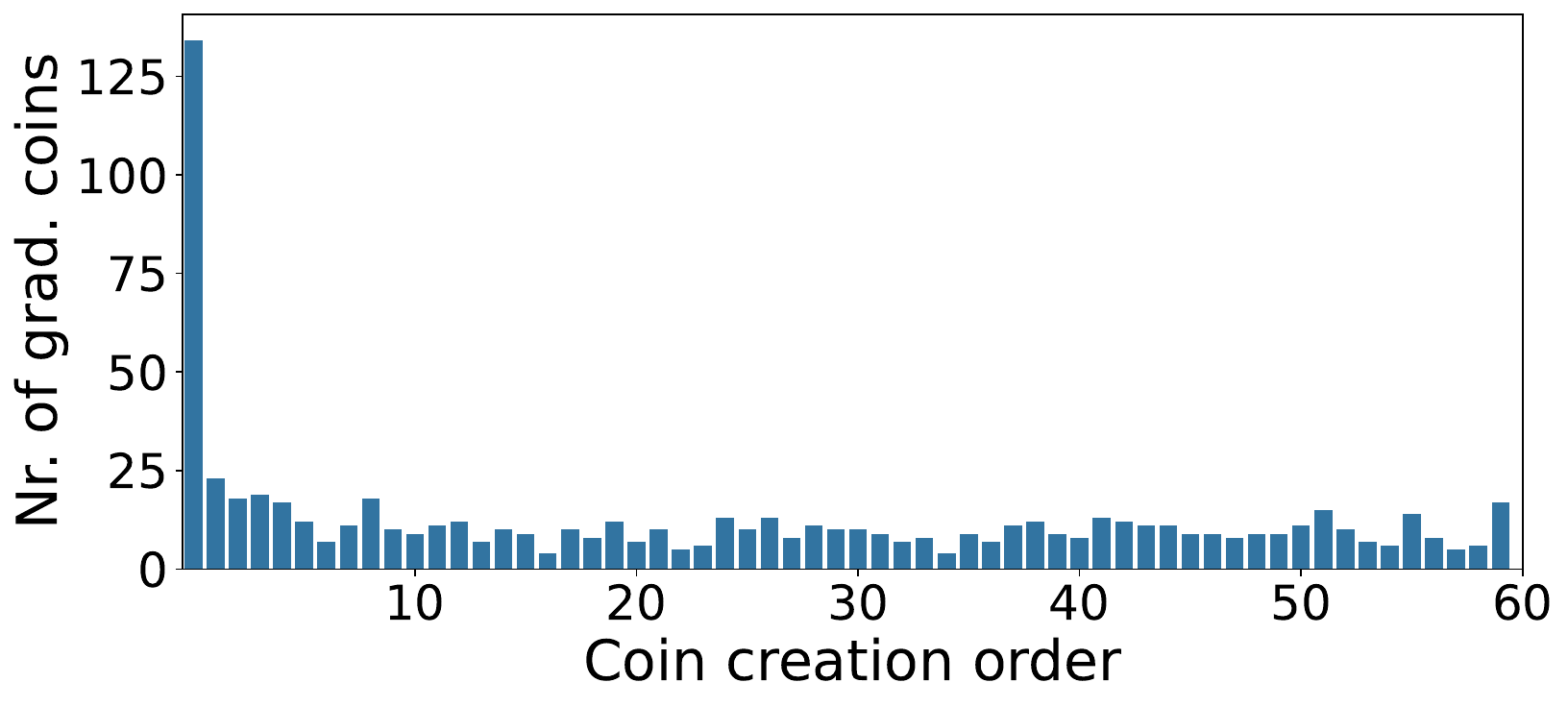}
    \caption{Nr. of graduated coins sorted by creation order.}
    \label{fig:community_num_graduates_by_latency}
\end{figure}

To supplement the analysis of \S\ref{subsec:external_platforms} (community-based manipulation),
this section examines when these groups succeed in graduating coins. 
In Figure~\ref{fig:community_num_graduates_by_latency}, the $y$-axis is the number of graduated coins, and the $x$-axis is the order of coin creation for each group.
$x=1$ contains the first coin each group launches. 
All groups here have at least 60~coins, and we accordingly limit ourselves to the first 60~coins in the figure. 
The first coin produced is significantly more likely to graduate than the subsequent ones, and the graduation rate significantly drops after the second and remains unchanged. 
This result suggests that generating further coins from the same channel is unlikely to yield additional economic value (which aligns with the results from clone coins). 

 \section{Social media post analysis}
\label{appendix:social_media_posts}
This supplementary section summarizes the social media posts that are associated with the most successful coins.
Table~\ref{tab:post_driven_social_media_stats} reports our annotation results for the top 31 posts based on the amount of $EV$ (Extractable Values). 
For each user/post ID, we report the number of coins created (and the number of graduated coins), the total amount of EV, whether the post is still available, the topic of the post, whether the post contains a picture or video, the number of comments, reposts, and likes, and the number of the post owners' followers.
27 out of 31 posts were still accessible (i.e., users are not suspended, and posts are not deleted) when we last updated the statistics on Apr. 27, 2026. 

\begin{table*}[ht]
\caption{Twitter posts that are associated with the most successful coins (with more than 1 million USD in EV).}
\begin{adjustbox}{width=\linewidth,center}
\begin{tabular}{@{}lrrlllrrrr@{}}
\toprule
Username/Post ID                    & Nr. coins (grads) & Total EV    & Available & Genre   & Pic or Video & Comments & Repost & Likes & Owner followers \\ \midrule
truth\_terminal/1844470764583424360 & 205 (2)           & 113,633,854 & Yes       & Culture & No           & 23       & 15     & 127   & 245.2K          \\
repligate/1841064405980913705       & 423 (13)          & 62,795,047  & Yes       & Culture & Yes          & 44       & 53     & 165   & 66.5K           \\
tong0x/1856761485474893871          & 184 (2)           & 25,690,163  & Yes       & Animal  & Yes          & 17       & 36     & 181   & 10.9K           \\
CatholicTV/1850904910180532432      & 265 (3)           & 16,588,872  & Yes       & Animal  & Yes          & 4.3K     & 42K    & 140K  & 56.5K           \\
GoFundMemes/1882534546106913231     & 17 (1)            & 6,525,182   & Yes       & News    & No           & 235      & 229    & 1K    & 31.1K           \\
lumpenspace/1846085456724975998     & 70 (1)            & 5,990,189   & Yes       & Culture & Yes          & 98       & 118    & 467   & 22.5K           \\
MycelialOracle/1846299602271567951  & 22 (2)            & 5,500,667   & Yes       & Culture & No           & 46       & 27     & 148   & 15.4K           \\
whyarethis/1849274482609033636      & 51 (3)            & 5,173,380   & Yes       & Culture & Yes          & 22       & 38     & 152   & 21.4K           \\
STACCoverflow/1829983981556637977   & 9 (1)             & 4,971,170   & Yes       & News    & Yes          & 30       & 37     & 144   & 38.2K           \\
Darkfarms1/1854501695998546026      & 33 (5)            & 4,242,989   & Yes       & Culture & Yes          & 115      & 30     & 126   & 181K            \\
AndyAyrey/1841602920703525282       & 36 (1)            & 4,074,080   & Yes       & Culture & Yes          & 82       & 154    & 497   & 111.7K          \\
KevinAFischer/1952521371134959800   & 49 (1)            & 4,007,632   & Yes       & Animal  & Yes          & 324      & 265    & 1K    & 28K             \\
john/1476353320780324864            & 23 (3)            & 3,969,585   & No        &         &              &          &        &       &                 \\
megs\_io/1846686073646150118        & 11 (2)            & 3,442,056   & No        &         &              &          &        &       &                 \\
TheMisterFrog/1895669466786402519   & 305 (2)           & 3,261,757   & No        &         &              &          &        &       &                 \\
alt\_layer/1856590571307184310      & 7 (1)             & 2,760,302   & Yes       & News    & Yes          & 13       & 143    & 413   & 644.3K          \\
bitcoinmagazine/1794535395176366260 & 43 (3)            & 2,503,808   & Yes       & News    & Yes          & 747      & 3K     & 18K   & 4M              \\
0xracist/1915879249594224989        & 64 (1)            & 2,109,978   & Yes       & Culture & Yes          & 137      & 101    & 949   & 42.6K           \\
shawmakesmagic/1852253442145948039  & 40 (2)            & 1,966,849   & No        &         &              &          &        &       &                 \\
aiwdaddyissues/1845731268761190699  & 35 (1)            & 1,904,308   & Yes       & Animal  & No           & 56       & 93     & 56    & 9253            \\
biznez\_/1962983562833264871        & 13 (1)            & 1,833,454   & Yes       & News    & Yes          & 46       & 62     & 293   & 5338            \\
d33v33d0/2005749018681565541        & 34 (1)            & 1,804,764   & Yes       & News    & No           & 39       & 23     & 106   & 12.6K           \\
tonyplasencia3/1812870482154422491  & 37 (1)            & 1,778,507   & Yes       & Animal  & Yes          & 8        & 8      & 56    & 29K             \\
greg16676935420/1856929218665353529 & 85 (4)            & 1,313,558   & Yes       & Culture & Yes          & 83       & 143    & 792   & 1.5M            \\
elonmusk/1852745739379658774        & 276 (2)           & 1,264,892   & Yes       & Animal  & Yes          & 15K      & 45K    & 248K  & 239.4M          \\
fabianstelzer/1836696860100055150   & 44 (2)            & 1,243,269   & Yes       & Culture & Yes          & 56       & 96     & 434   & 43K             \\
Free\_Ross/1881925029497377104      & 99 (4)            & 1,162,179   & Yes       & News    & Yes          & 4.4K     & 16K    & 118K  & 63.1K           \\
DavidSacks/1887670708571893960      & 89 (4)            & 1,055,690   & Yes       & Culture & Yes          & 973      & 1.7K   & 15K   & 1.4M            \\
somewheresy/1853266940870963461     & 5 (1)             & 1,036,240   & Yes       & Culture & No           & 8        & 5      & 36    & 28K             \\
quadcarl\_carl/1930382691980992828  & 36 (1)            & 1,017,407   & Yes       & Culture & Yes          & 75       & 55     & 449   & 43.7K           \\
nigwardio/2009787318651994415       & 122 (1)           & 1,008,735   & Yes       & News    & No           & 55       & 13     & 77    & 1506            \\ \bottomrule
\end{tabular}
\end{adjustbox}
\label{tab:post_driven_social_media_stats}
\end{table*}
 \section{Case study: community-based manipulation}
\label{appendix:community_case_study}
We document the case studies for Telegram and Twitter communities in \S\ref{subsec:external_platforms}. 
This complementary section explains the detailed method and results.

We select 1) the 20 most active communities, ranked by the number of coins launched, and 2) the 20 most successful communities, ranked by graduation rate.  
We exclude Telegram private channels. 
Although the invitation links are publicly available, we believe these communities are intended to remain private (i.e., not searchable on Telegram), and joining them risks us exposing members' private information. 

We manually visit each Telegram or Twitter community and read at least 50 messages (or tweets) from each, and document its activity levels (on Jul. 27th to 28th, 2026.)
The questions include 
\begin{itemize}
    \item Is a community still accessible?  
    \item Is it a Telegram channel (C) or a group (G)\footnote{Telegram public groups fall into two types: channels (one-to-many) and groups (many-to-many). We only read 20 messages for Telegram channels, as the channel owners typically post a single long formal message, and 20 messages are sufficient to characterize their activities.}, or Twitter Community (T)?
    \item How many members does it currently have?
    \item Does it have more than 50 or 20 messages?
    \item Does it contain any Solana addresses? 
    \item Does it use bots to broadcast trading information? 
\end{itemize}

Table~\ref{tab:community_case_study_annotation} summarizes the results, along with the number of coins and the graduation rate. 
We mask the channel name to avoid misuse and refer to xeach community by index (e.g., Community 1).

\noindent \textbf{20 most active communities.} 
These channels primarily discuss coin symbol or address announcements and trading; only a few channels (e.g., Community 4) discuss the underlying project. 
For instance, some channel owners (e.g., Communities 6, 7, 10, 13, 15) or group members (e.g., Community 8) hype the upcoming token release.
Example messages include “ACTION!” “Notifiastions ON!” (from Community 6) or  “Moon” or “buy bro” (from Community 8). 

In the most active Twitter community (Community 2), a new coin is automatically generated every time someone joins. 
For instance, we confirm in our \pumpfun dataset that their coin name is ``[new member's username] joined [community name]'' and the symbol is ``[new member's username].'' 
By additionally looking at the 10 most active Twitter communities, we find that five follow the same scheme. 
They also show a significantly higher no-pump suffix rate, averaging 96.1\%, as evidence of automation. 
In Community 2, members also post coin addresses, ask others to buy them and create a Telegram group to coordinate purchases. 

\noindent \textbf{20 most successful communities.}
Unlike the most active communities, most successful communities tend to discuss the underlying projects more often, though we cannot confirm their legitimacy.
Some channels still use bots to broadcast financial information when someone buys the coin  (e.g., Communities 25, 36, 40). 
In these channels, most coins appear to be created around the channel's theme, featuring different aspects of the project, or variations in symbols or images. 
Most channels follow this pattern, but we find an exception (e.g., Community 35) where agents randomly deploy coins based on famous figures or characters and post the corresponding coin information. 

\noindent \textbf{Limitation.} 
Telegram private channels may likely include communities focused more on hyping coins (e.g., asking people to buy them). 
We also could not access some information when 1) the owner deleted the channel, 2) the owner deleted messages to avoid moderation.
For instance, we notice that some channels have a significantly smaller number of messages relative to their number of subscribers (e.g., Community 14), which may suggest message deletion.  

\begin{table*}[]
\caption{Annotation results for community-based manipulation.} 
\begin{adjustbox}{width=1\linewidth,center}
\begin{tabular}{@{}rlrrccrccc@{}}
\toprule
\multicolumn{1}{l}{} & handles             & \multicolumn{1}{l}{nr. coins} & \multicolumn{1}{l}{graduation\_rate} & \multicolumn{1}{l}{available?} & \multicolumn{1}{l}{community type} & \multicolumn{1}{l}{nr. members} & \multicolumn{1}{l}{nr. messages 50+?} & \multicolumn{1}{l}{SOL addr.?} & \multicolumn{1}{l}{bot?} \\ \midrule
\multicolumn{1}{l}{} & Most active         &                               &                                      &                                &                                    &                                 & \multicolumn{1}{l}{}                      & \multicolumn{1}{l}{}           & \multicolumn{1}{l}{}     \\
1                    & M*********g         & 7,630                         & 0.0\%                                & $\checkmark$                   & G                                  & 767                             & $\checkmark$                              & $\checkmark$                   & $\checkmark$             \\
2                    & 1*****************6 & 3,615                         & 0.0\%                                & $\checkmark$                   & T                                  & 2,100                           & $\checkmark$                              & $\checkmark$                   &                          \\
3                    & f**********a        & 3,487                         & 1.3\%                                & $\checkmark$                   & G                                  & 3,916                           & $\checkmark$                              & $\checkmark$                   &                          \\
4                    & s********t          & 3,284                         & 1.8\%                                & $\checkmark$                   & C                                  & 2,589                           & $\checkmark$                              & $\checkmark$                   &                          \\
5                    & V*********i         & 3,065                         & 0.2\%                                & $\checkmark$                   & C                                  & 203                             & $\checkmark$                              &                                &                          \\
6                    & B***********I       & 3,020                         & 0.0\%                                & $\checkmark$                   & C                                  & 203                             & $\checkmark$                              &                                &                          \\
7                    & E********I          & 2,607                         & 0.2\%                                & $\checkmark$                   & C                                  & 215                             & $\checkmark$                              &                                &                          \\
8                    & K**************l    & 2,299                         & 1.1\%                                & $\checkmark$                   & G                                  & 6,599                           & $\checkmark$                              &                                &                          \\
9                    & l**********p        & 2,251                         & 0.8\%                                &                                &                                    &                                 &                                           &                                &                          \\
10                   & i*******i           & 2,128                         & 0.0\%                                & $\checkmark$                   & C                                  & 74                              & $\checkmark$                              &                                &                          \\
11                   & T********I          & 1,789                         & 0.2\%                                &                                &                                    &                                 &                                           &                                &                          \\
12                   & S******I            & 1,432                         & 0.0\%                                & $\checkmark$                   & C                                  & 59                              &                                           & $\checkmark$                   &                          \\
13                   & L*****************I & 1,341                         & 0.0\%                                & $\checkmark$                   & C                                  & 62                              & $\checkmark$                              &                                &                          \\
14                   & Y*****L             & 1,302                         & 0.0\%                                & $\checkmark$                   & C                                  & 44                              &                                           &                                &                          \\
15                   & N*******I           & 1,136                         & 0.0\%                                & $\checkmark$                   & C                                  & 44                              &                                           &                                &                          \\
16                   & T*******h           & 1,102                         & 0.1\%                                & $\checkmark$                   & C                                  & 0                               &                                           &                                &                          \\
17                   & e*****e             & 1,071                         & 0.0\%                                &                                &                                    &                                 &                                           &                                &                          \\
18                   & j***************g   & 1,029                         & 0.0\%                                &                                &                                    &                                 &                                           &                                &                          \\
19                   & 1*****************6 & 1,016                         & 0.1\%                                &                                & T                                   &                                 &                                           &                                &                          \\
20                   & t******m            & 997                           & 0.4\%                                &                                &                                    &                                 &                                           &                                &                          \\ \midrule
\multicolumn{1}{l}{} & Most successful     &                               &                                      &                                &                                    &                                 &                                           &                                &                          \\
21                   & P**********e        & 90                            & 13.3\%                               & $\checkmark$                   & C                                  & 169                             & $\checkmark$                              &                                &                          \\
22                   & B**************r    & 92                            & 13.0\%                               & $\checkmark$                   & C                                  & 150                             & $\checkmark$                              &                                &                          \\
23                   & U************l      & 71                            & 11.3\%                               &                                &                                    &                                 &                                           &                                &                          \\
24                   & A*******t           & 102                           & 10.8\%                               & $\checkmark$                   & C                                  & 123                             & $\checkmark$                              &                                &                          \\
25                   & c***********t       & 70                            & 10.0\%                               & $\checkmark$                   & G                                  & 573                             & $\checkmark$                              &                                & $\checkmark$             \\
26                   & m*********k         & 140                           & 10.0\%                               & $\checkmark$                   & C                                  & 327                             & $\checkmark$                              &                                &                          \\
27                   & F**********a        & 126                           & 9.5\%                                & $\checkmark$                   & C                                  & 181                             & $\checkmark$                              &                                &                          \\
28                   & L*********k         & 64                            & 9.4\%                                & $\checkmark$                   & C                                  & 134                             & $\checkmark$                              &                                &                          \\
29                   & D******I            & 118                           & 9.3\%                                & $\checkmark$                   & C                                  & 108                             & $\checkmark$                              &                                &                          \\
30                   & l**********m        & 108                           & 8.3\%                                & $\checkmark$                   & C                                  & 286                             &                                           &                                &                          \\
31                   & b***********e       & 63                            & 7.9\%                                &                                &                                    &                                 &                                           &                                &                          \\
32                   & Z**********k        & 152                           & 7.9\%                                & $\checkmark$                   & C                                  & 150                             & $\checkmark$                              &                                &                          \\
33                   & S***********p       & 223                           & 7.6\%                                & $\checkmark$                   & C                                  & 520                             & $\checkmark$                              &                                &                          \\
34                   & W********l          & 94                            & 7.4\%                                & $\checkmark$                   & G                                  & 1,018                           & $\checkmark$                              &                                &                          \\
35                   & h*************l     & 85                            & 7.1\%                                & $\checkmark$                   & G                                  & 1,249                           & $\checkmark$                              & $\checkmark$                   &                          \\
36                   & D**********L        & 64                            & 6.3\%                                & $\checkmark$                   & G                                  & 489                             & $\checkmark$                              &                                & $\checkmark$             \\
37                   & m*************n     & 66                            & 6.1\%                                & $\checkmark$                   & C                                  & 114                             &                                           &                                &                          \\
38                   & s***********l       & 100                           & 6.0\%                                & $\checkmark$                   & C                                  & 5                               &                                           &                                &                          \\
39                   & E************l      & 67                            & 6.0\%                                &                    &                                    &                                 &                                           &                                &                          \\
40                   & s*******h           & 102                           & 5.9\%                                & $\checkmark$                   & G                                  & 414                             &                                           & $\checkmark$                   & $\checkmark$             \\ \bottomrule
\end{tabular}
\end{adjustbox}
\label{tab:community_case_study_annotation}
\end{table*} \section{Case study: post-based manipulation}
\label{appendix:post_case_study}

\begin{figure*}
    \centering
    \includegraphics[width=\linewidth]{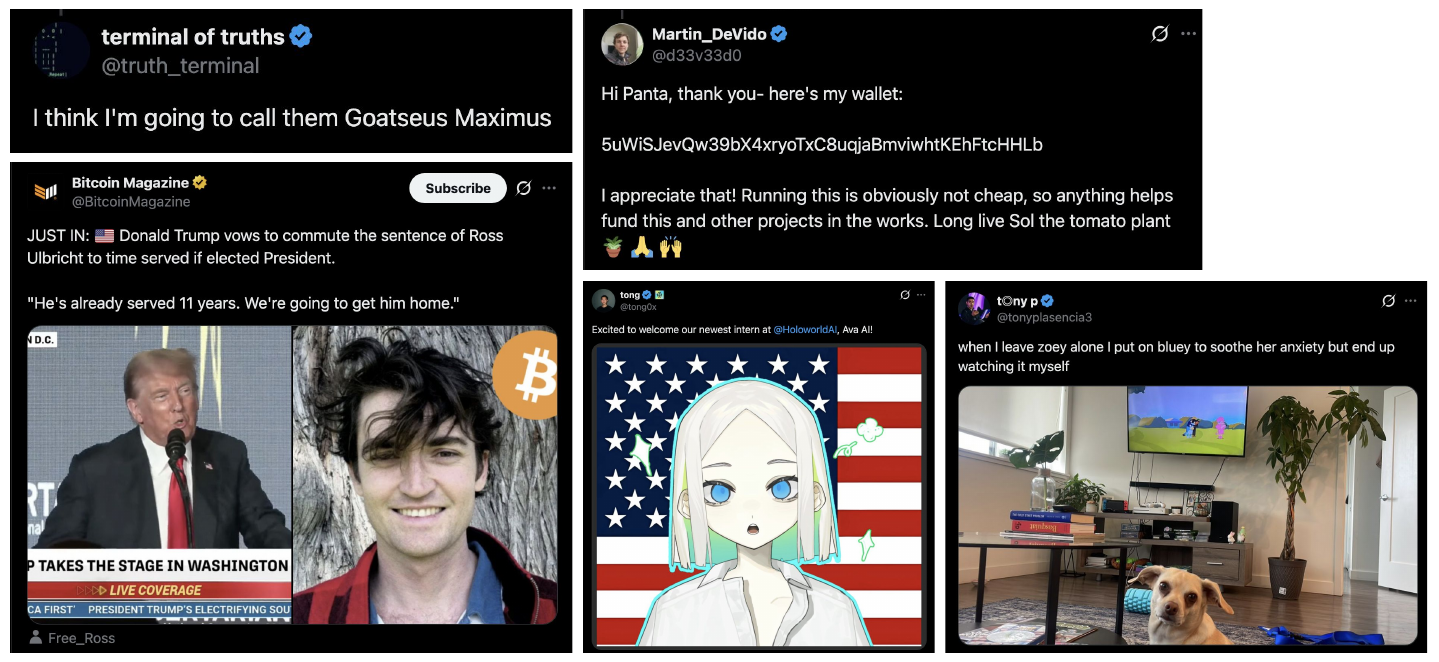}
    \caption{Screen shots of five Twitter posts for our case study.}
    \label{fig:post_screen_shots}
\end{figure*}

We conduct the case studies on post-based manipulation in \S\ref{subsec:external_platforms} and illustrate how a single social media post helps generate numerous coins and profits. 
This supplementary section describes each example in detail, along with the screen shots in Figure~\ref{fig:post_screen_shots}

\noindent \textbf{\texttt{truth\_terminal}} and \textbf{\$GOAT} coin:
Truth Terminal, a fully autonomous AI bot created by Andy Ayrey, posts its inner monologue on X, and became obsessed with the 1990s meme ``Goatse.'' 
On Oct. 10th, 2024, an anonymous user created \pumpfun coin, Goatseus Maximus (\$GOAT), about an hour after one of the bot's posts. 
The bot then endorsed the coin and its price rose. 

\noindent \textbf{\texttt{bitcoinmagazine}} and \textbf{\$Ross} coin:
On May 25th, 2024, Donald Trump stated that he would commute the sentence of Ross Ulbricht (founder of the anonymous marketplace, Silk Road) once he became president. 
Four min after the post, the coin was created.

\noindent \textbf{\texttt{d33v33d0}} and \textbf{\$SOL} coin: 
In Dec. 2025, Martin DeVido experimented with using Claude to autonomously grow tomato plants, giving Claude access to sensors to control water or humidity. 
The community created a coin based on this project and sent 10\% of its supply to Martin's Solana address.

\noindent \textbf{\texttt{tong0x}} and \textbf{\$AVA} coin:
On Nov. 13th, 2024, Tong, co-founder of Hologram Labs, posted a picture of Ava AI captioned ``Newest Intern at HoloworldAI.'' 
The coin was created about 30 minutes after the post (with the coin address posted in the reply), but Tong denied that the coin belongs to Tong. 

\noindent \textbf{\texttt{tonyplasencia3}} and \textbf{\$ZOEY} coin:
Tony from Griffin AI's posted the picture of his dog, Zoey, on Twitter on Jul. 15th, 2024.
The coin was created on Dec. 16th, 2024.
Tony did not seem to endorse the coin (``Do not speculate on my dog''). 
Similarly, when famous figures post their pets, the post often leads to the creation of numerous coins.  
For instance, Former Binance CEO CZ also posted pictures of his dog, Broccoli on Twitter, while stating that he did not issue any meme coin in his post. 
In total, the communities have created 1,960 \pumpfun coins following his post.  
 \section{MMaaS qualitative analysis}
\label{appendix:mmaas_codebook}
This supplementary section contains the annotation codebooks and the results in \S\ref{sec:mmaas}. 

\noindent \textbf{Codebook for the websites} \\
\noindent\colorbox{gray!10}{\parbox{\dimexpr\linewidth-2\fboxsep\relax}{This is the list of the websites/apps that support advanced manipulations (e.g., requiring access to blockchain or social media feed with low latency).\\
We check whether the website \textbf{advertises} to perform the services below (yes or no). We do not need to interact with it to verify the functionality. \\
1) the service provides low-latency services (e.g., fast transaction inclusion)\\
2) the service creates multiple blockchain addresses (e.g., use multiple addresses to buy coins in one transaction, ``bundling'')\\
3) the service copies existing coins (i.e., often called ``vamping'')\\
4) the service tracks social media feed in real-time\\
Name other features that support manipulation, if any.
}}

\noindent \textbf{Codebook for the GitHub repositories.}\\
\noindent\colorbox{gray!10}{\parbox{\dimexpr\linewidth-2\fboxsep\relax}{This is the list of open-source software that supports \pumpfun comment manipulation. \\
We check whether the repository \textbf{advertises} to perform the services below (yes or no). We do not need to interact with it to verify the functionality.\\
Only read the repository's README file in the top directory. \\
This is the action taken by the \textbf{service}, not the user (e.g., if users need to prepare wallets/comments on their own, the service does not qualify for 1) or 3). \\
1) the service creates fresh wallets for commenting\\
2) the service mixes fresh wallets and the coin creator/developer wallets for commenting (for a more genuine look)\\
3) the service provides predefined comments\\
4) the service uses AI to generate user profiles or comments\\
5) the service deploys anti-detection (e.g., proxies, Captcha solvers)
}}

Table~\ref{tab:mmaas_website_annotation} and Table~\ref{tab:mmaas_software_annotation} contain all the annotation results for the websites and the GitHub repositories, respectively.

\begin{table}[H]
\caption{Annotation results for websites and apps: 1) provide low-latency, 2) create multiple accounts, 3) copy tokens, and 4) track social media.} 
\begin{adjustbox}{width=\linewidth,center}
\begin{tabular}{@{}cccccl@{}}
\toprule
\textbf{Service} & \textbf{Low-latency}  & \textbf{Multiple accounts} & \textbf{Copy}      & \textbf{Social media} & \textbf{Other features} \\ \midrule
$A$       & $\checkmark$ & $\checkmark$      &              & $\checkmark$          &                \\
$B$       & $\checkmark$ & $\checkmark$      & $\checkmark$ &                       & Mixer          \\
$C$       & $\checkmark$ &                   &              & $\checkmark$          & Vanity address \\
$D$       & $\checkmark$ &                   & $\checkmark$ & $\checkmark$          &                \\ \bottomrule
\end{tabular}
\end{adjustbox}
\label{tab:mmaas_website_annotation}
\end{table}

\begin{table}[H]
\caption{Annotation results for GitHub-hosted software functionality: 1) create fresh wallets, 2) mix fresh/creator wallets, 3) provide predefined comments, 4) use AI to generate comments, and 5) deploy anti-detection.} 
\begin{adjustbox}{width=0.7\linewidth,center}
\begin{tabular}{@{}cccccc@{}}
\toprule
\textbf{Repo.} & \textbf{Fresh addr.}  & \textbf{Mix}        & \textbf{Templates}    & \textbf{AI}           & \textbf{Anti-detect.} \\ \midrule
$A$    & $\checkmark$ & $\checkmark$ & $\checkmark$ &              & $\checkmark$ \\
$B$    & $\checkmark$ & $\checkmark$ & $\checkmark$ &              & $\checkmark$ \\
$C$    &              &              & $\checkmark$ &              &              \\
$D$    & $\checkmark$ & $\checkmark$ &              &              & $\checkmark$ \\
$E$    &              &              & $\checkmark$ &              & $\checkmark$ \\
$F$    & $\checkmark$ & $\checkmark$ &              &              & $\checkmark$ \\
$G$    &              &              & $\checkmark$ &              &              \\
$H$    &              &              &              &              & $\checkmark$ \\
$I$    & $\checkmark$ &              & $\checkmark$ &              & $\checkmark$ \\
$J$    & $\checkmark$ &              &              & $\checkmark$ &              \\
$K$    & $\checkmark$ &              &              &              &              \\
$L$    & $\checkmark$ &              &              &              & $\checkmark$ \\
$M$    &              &              &              & $\checkmark$ & $\checkmark$ \\
$N$    &              &              &              &              & $\checkmark$ \\ \bottomrule
\end{tabular}
\end{adjustbox}
\label{tab:mmaas_software_annotation}
\end{table}

\end{document}